%% file: HaloModel.WDM.4.tex
\documentclass[10pt,twocolumn,preprintnumbers,amsmath,amssymb]{revtex4}

\usepackage{natbib}
\usepackage{graphicx}  
\usepackage{dcolumn}   
\usepackage{bm}        

\input{defs.tex}

\def\s{{\rm s}}
\def\h{{\rm h}}
\def\ss{{\rm ss}}
\def\hh{{\rm hh}}

\def\1H{{\rm 1H}}
\def\2H{{\rm 2H}}

\def\WDM{{\rm WDM}}
\def\CDM{{\rm CDM}}
\def\keV{{\rm keV}}
\def\kms{{\rm km\,s^{-1}}}

\def\cut{{\rm cut}}

\begin{document}


\title{Testing the Warm Dark Matter paradigm with large-scale structures}


\author{Robert E. Smith$^{1,2}$ \& Katarina Markovic$^{3,4}$}

\affiliation{
\vspace{0.2cm}
(1) Institute for Theoretical Physics, University of Zurich, Zurich CH 8037, Switzerland\\
(2) Argelander-Institute for Astronomy, Auf dem H\"ugel 71, D-53121 Bonn, Germany\\
(3) University Observatory Munich, Ludwig-Maximillian University, Scheinerstr. 1, 81679 Munich, Germany\\
(4) Excellence Cluster Universe, Boltzmannstr. 2, 85748 Garching, Germany} 
\email{res@physik.unizh.ch, markovic@usm.lmu.de}


\begin{abstract}
  We explore the impact of a $\Lambda$WDM cosmological scenario on the
  clustering properties large-scale structure in the Universe. We do
  this by extending the halo model. The new development is that we
  consider two components to the mass density: one arising from mass
  in collapsed haloes, and the second from a smooth component of
  uncollapsed mass. Assuming that the nonlinear clustering of dark
  matter haloes can be understood, then from conservation arguments
  one can precisely calculate the clustering properties of the smooth
  component and its cross-correlation with haloes. We then explore how
  the three main ingredients of the halo calculations, the halo mass
  function, bias and density profiles are affected by WDM. We show
  that, relative to CDM, the halo mass function is suppressed by
  $~50\%$, for masses $\sim100$ times the free-streaming mass-scale
  $M_{\rm fs}$.  Consequently, the bias of low mass haloes can be
  boosted by as much as $\sim20\%$ for 0.25\,keV WDM particles. Core
  densities of haloes will also be suppressed relative to the CDM
  case. We also examine the impact of relic thermal velocities on the
  density profiles, and find that these effects are constrained to
  scales $r<1\kpc$, and hence of little importance for dark matter
  tests, owing to uncertainties in the baryonic physics. We use our
  modified halo model to calculate the non-linear matter power
  spectrum, and find that there is significant small-scale power in
  the model. However, relative to the CDM case the power is
  suppressed. The amount of suppression depends on the mass of the WDM
  particle, but can be of order 10\% at $k\sim1\kMpc$ for particles of
  mass 0.25\,keV. We then calculate the expected signal and noise that
  our set of $\Lambda$WDM models would give for a future weak lensing
  mission. We show that the models should in principle be separable at
  high significance. Finally, using the Fisher matrix formalism we
  forecast the limit on the WDM particle mass for a future full-sky
  weak lensing mission like Euclid or LSST.  With Planck priors and
  using only multipoles $l<5000$, we find that a lower limit of
  $2.6\,\keV$ should be easily achievable.
\end{abstract}




\maketitle


\section{Introduction}\label{sec:intro}

Within the last two decades cosmology has experienced a major
observational revolution. The measurement of the Cosmic Microwave
Background (CMB) anisotropies, through such surveys as COBE and WMAP,
and the mapping of the large-scale structures (LSS), through galaxy
redshift surveys like 2dFGRS and SDSS, have had a great impact on our
knowledge of the Universe. In particular, it has led to the
establishment of the 'standard model' for cosmology: $\Lambda$CDM
model \citep{Spergeletal2007} -- meaning a Universe whose mass density
is dominated by collisionless Cold Dark Matter (CDM) relic particles,
but whose total energy density today is dominated by the constant
energy of the vacuum $\Lambda$, which is driving the current
accelerated expansion of the Universe.  However, fundamental questions
remain unanswered: What is the true physical nature of the cold dark
matter? and what is the true origin of the accelerated expansion?  In
this paper we focus on developing tests for the dark matter model
based upon the influence that different particle candidates have upon
the statistical properties of the LSSs.

There is a large body of indirect astrophysical evidence that strongly
supports the CDM hypothesis
\citep{Bertoneetal2004,DiemandMoore2009,Bertone2010} -- i.e. a heavy,
cold thermal relic particle, e.g. weakly interacting massive
particles, etc., that decoupled from normal matter very early in the
history of the Universe.  However, despite the many successes of
$\Lambda$CDM, there is also a growing body of evidence in conflict
with parts of the hypothesis. Firstly, hi-resolution simulations of
CDM predict several hundred dwarf galaxy halo satellites within the
Local Group, whereas the observed abundance of dwarf galaxies is a
factor of 10 lower
\citep{Mooreetal1999b,Klypinetal1999,PeeblesNusser2010,Diemandetal2007,Springeletal2008},
although counting and weighing local group galaxies is not easy
\citep{Zavalaetal2009,MaccioFontanot2010}. Secondly, the inner density
profiles of haloes in CDM simulations are cuspy, with the inner
density falling as $\rho(r)\propto r^{-1}$
\citep{Navarroetal1997,Mooreetal1999c,Ghignaetal2000,Diemandetal2004,Stadeletal2009},
whereas the density profiles inferred from galaxy rotation curves are
significantly shallower \citep{Mooreetal1999c} (and for more recent
studies see
\citep{Swatersetal2003,Saluccietal2007,deBloketal2008,Gentileetal2009}
and references there in).  Thirdly the observed number of dwarf
galaxies in the voids is far smaller than expected from CDM
\citep{PeeblesNusser2010}.

Over the past decade much work has been invested in attempting to
understand these discrepancies. Some of these problems may be
reconciled through more careful treatment of baryonic physics and
galaxy formation.  More exotic explanations for some of these problems
would require modifications to gravity or scalar fields that operate
only in the dark matter sector
\citep{PeeblesNusser2010,Kroupaetal2010}. Perhaps a less dramatic
explanation might be the modification of the physical properties of
the dark matter particle.

Over the past decade, several modifications to the CDM hypothesis have
been forwarded \citep[see for example][]{HoganDalcanton2000}.  In this
work we focus on Warm Dark Matter (WDM) in a low-density $\Lambda$
dominated universe (hereafter $\Lambda$WDM). In this scenario, the
dark particle is considered to be lighter than its CDM counterpart,
and retains some stochastic primordial thermal velocity, whose
momentum distribution obeys Fermi-Dirac statistics
\citep{Bodeetal2001,Boyanovsky2010}.  Two potential candidates are the
gravitino \citep{Ellisetal1984} and a sterile neutrino
\citep{DodelsonWidrow1994}, both of which require extensions of the
standard model of particle physics
\citep{Bodeetal2001,Boyarskyetal2009a}. The main cosmological effect
of WDM is that, at early times while it remains relativistic, density
fluctuations are suppressed on scales of order the Hubble horizon at
that time -- and particles are said to free-stream out of
perturbations. In fact, until matter-radiation equality, the WDM
particles can also still travel a comparable distance whilst
non-relativistic. Consequently, the matter power spectrum becomes
suppressed on small scales. At later times the particles cool with the
expansion and at the present time essentially behave like CDM --
modulo any remaining primordial velocity dispersion \citep[for more
  details see][]{DodelsonWidrow1994,Bodeetal2001,Boyanovsky2010}.

Past research on the impact of the WDM hypothesis on the phenomenology
of cosmic structures has primarily focused on small-scale phenomena
\citep{Colinetal2000,WhiteCroft2000,Avila-Reeseetal2001,Bullocketal2002,Colinetal2008,Zavalaetal2009},
although there are some notable studies of large-scale structures
\citep{Colombietal1996,Bodeetal2001}. In short, these results show
that halo abundances for masses below the free-streaming mass-scale
are suppressed relative to CDM.  The amount of suppression is still a
point of debate: initial studies, suggested that low mass haloes might
form through the fragmentation of larger perturbations, in particular
along filaments \citep{Bodeetal2001}. However, recent numerical
studies \citep{WangWhite2007,PolisenskyRicotti2010} have cast doubt on
this as a viable mechanism, since the abundance of structure forming
below the free-streaming mass-scale appears to be correlated with the
initial particle grid and inter-particle separation. Some simulation
results have also claimed that the density profiles of WDM haloes are
shallower in the inner core regions \citep{Colinetal2008}. Further,
whilst the linear matter power spectrum is suppressed, the nonlinear
one regenerates a high $k$-tail \citep{WhiteCroft2000}.

Observational constraints suggest that sterile neutrinos can not be
the dark matter: the Lyman alpha forest bounds are $m_{\nu_s}>8\keV$
\citep{Seljaketal2006b,Boyarskyetal2009a}, whilst those from the X-ray
background are $m_{\nu_s}<4\keV$ \citep{Boyarskyetal2008}.  However a
more recent assessment has suggested that a better motivated particle
physics model for the production of the sterile neutrino may evade
these constraints, with the new bound $m_{\nu_s}\gtrsim2\keV$
\citep{Boyarskyetal2009b}. It therefore seems that independent methods
for constraining the properties of WDM from cosmology would be
valuable.

In a recent paper, \citep{Markovicetal2010}, it was proposed that the
$\Lambda$WDM scenario could be tested through weak lensing by LSS. The
advantage of such a probe is that it is only sensitive to the total
mass distribution along the line of sight. However, to obtain
constraints on the WDM particle mass, an accurate model for the
nonlinear matter clustering is required.  \citet{Markovicetal2010}
attempted to construct this in two ways, firstly through use of the
halo model and secondly through adapting a fitting formula for CDM
power spectra. There appeared to be some discrepancy between the
predictions of the two approaches. Furthermore, their halo model
calculation suggested that the bias of high mass haloes was
significantly larger than for the case of CDM. One of the aims of this
paper is to construct an improved halo model calculation for
application to this problem.

The paper is broken down as follows. In \S\ref{sec:WDMtheory} we
briefly review the impact of WDM on the linear theory evolution of
structure. In \S\ref{sec:halomodel} we describe our new halo model
approach. In \S\ref{sec:ingredients} we discuss the ingredients of the
halo model in the context of WDM. In \S\ref{sec:results} we show our
results for the clustering statistics and the weak lensing
observables.  In \S\ref{sec:fisher} we forecast a limit on the
particle mass, obtainable from a future full-sky weak lensing survey.
In \S\ref{sec:conclusions} we summarize our findings and conclude.


\section{Theoretical Background}\label{sec:WDMtheory}


\begin{figure}
\centering{
  \includegraphics[width=8cm,clip=]{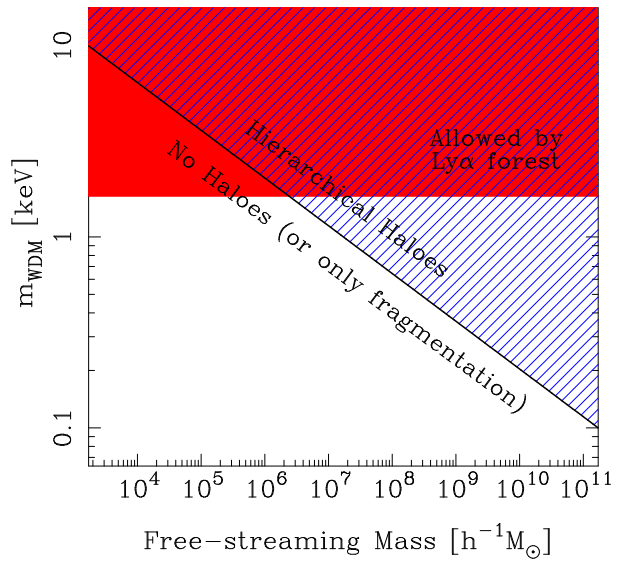}}
\caption{\small{Free-streaming mass-scale ($M_{\rm fs}$) as a function
    of the mass of the WDM particle mass $m_{\WDM}$. Haloes that have
    masses $M>M_{\rm fs}$ may form hierarchically (hatched
    region). Haloes with $M<M_{\rm fs}$ (empty region) will have their
    peaks in the primordial density field erased and so may not form
    hierarchically. Haloes in this region may possibly form through
    fragmentation. The solid red band shows the $m_{\WDM}$ allowed by
    the Lyman alpha forest \citep{Boyarskyetal2009a} (note that we
    have rescaled $m_{\nu_s}\rightarrow m_{\WDM}$ using
    \Eqn{eq:massrescale}).
 \label{fig:massFS}}}
\end{figure}


One of the major differences between structure formation in the
$\Lambda$CDM and $\Lambda$WDM scenarios, is that whilst the dark
matter particle is relativistic it free-streams through the Universe
and so suppresses the growth of density perturbations on scales
smaller than the free-streaming length
\citep{Bondetal1980,Colombietal1996,Bodeetal2001,Vieletal2005}
\footnote{Note that the free-streaming mass scale is not the same as
  the Jeans mass. In WDM the Jeans mass is set by the effective
  pressure of the relic thermal motions of the particles, whereas the
  free-streaming mass is the linear mass scale at which significant
  departures from a CDM model are expected.  At early times when the
  particles are relativistic these two mass scales can coincide or are
  proportional. However, at late times, the Jeans length drops
  significantly below the free-streaming mass scale $M_{\rm fs}$, and
  if the relic velocities are absent, then it is zero.}.
The lighter the WDM particle the longer it remains relativistic and
hence the larger the free-streaming scale. Following convention, we
shall define the free-streaming scale as the half-wavelength of the
mode for which the linear perturbation amplitude is suppressed by a
factor of 2 \citep{Avila-Reeseetal2001,Bodeetal2001,Vieletal2005}.
Fitting to the solutions of integrations of the collisionless
Einstein-Boltzmann equations through the epoch of recombination, sets
the free-streaming scale to be \citep{ZentnerBullock2003}:
\be \lambda_{\rm fs} \approx 0.11
\left[\frac{\Omega_{\WDM}h^2}{0.15}\right]^{1/3}
\left[\frac{m_{\WDM}}{\keV} \right]^{-4/3} {\rm Mpc}\ ,\ee
where $\Omega_{\WDM}$ is the total energy density in WDM, relative to
the critical density for collapse, and $m_{\WDM}$ is the mass of the
dark matter particle in keV.  From the scale $\lambda_{\rm fs}$ we may
infer a putative halo mass whose formation is suppressed by the
free-streaming. In the initial conditions this is given by
\citep{Avila-Reeseetal2001}:
\be M_{\rm fs}=\frac{4}{3}\pi\left(\frac{\lambda_{\rm fs}}{2}\right)^3\rhob \ .\ee

\begin{figure}
\centering{
  \includegraphics[width=8.0cm,clip=]{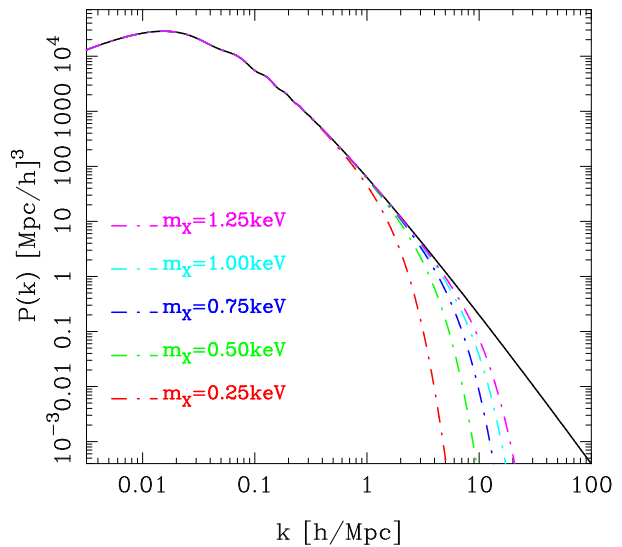}}
\caption{\small{Linear mass power spectra as a function of wavenumber
in the WDM and CDM scenarios. The solid black line shows results for
CDM. The dot-dashed lines represent the WDM power spectra for the
cases where \mbox{$m_{\WDM}\equiv m_X\in\{0.25,\,0.5,\,0.75,\,1.0,\,1.25\}\,{\rm
keV}$}. The lighter the WDM particle the more the power is damped on
small scales.}\label{fig:powerWDM}}
\end{figure}

In \Fig{fig:massFS} we show the relation between the free-streaming
mass-scale and the mass of the WDM particle candidate for our adopted
cosmological model. Haloes that would have masses below the
free-streaming mass-scale will have their peaks in the primordial
density field erased. The impact of the free-streaming of the WDM
particles on the linear growth of structure can be further described
by the matter transfer function. This can be represented
\citep{Bodeetal2001,Markovicetal2010},
\be 
T(k)\equiv \left[\frac{P^{\WDM}_{\rm Lin}}{P^{\CDM}_{\rm Lin}}\right]^{1/2} =
\left[1+(\alpha k)^{2\mu}\right]^{-5/\mu}\ ,
\ee
where $P^{\WDM}_{\rm Lin}$ and $P^{\CDM}_{\rm Lin}$ are the linear
mass power spectra in the WDM and CDM models respectively. Note that
in order to avoid cumbersome notation, we shall adopt the convention
$P_{\rm Lin}\equiv P^{\WDM}_{\rm Lin}$, likewise $P_{\rm NL}\equiv
P_{\NL}^{\WDM}$ for the nonlinear power spectrum in the WDM
model. Fitting to results of full Einstein-Boltzmann integrations
gives $\mu=1.12$, and $\alpha$ of the order \citep{Vieletal2005}:
\be 
\alpha = 0.049 \left[\frac{m_{\WDM}}{\keV}\right]^{-1.11}\!\!
\left[\frac{\Omega_{\WDM}}{0.25}\right]^{0.11}\!\left[\frac{h}{0.7}\right]^{1.22}
\!\!\!\Mpc \ .
\ee
In \Fig{fig:powerWDM} we show how variations in the mass of the WDM
particles effects the linear matter power spectrum. Clearly, the lighter
the WDM particle the more the power is damped on small scales.

Note that in the above we are assuming that the WDM particle is fully
thermalized, i.e. the gravitino. Following \citep{Vieletal2005}, this
can be related to the mass for the sterile neutrino through the
fitting formula:
\be 
m_{\nu_s}=4.43 \keV \left(\frac{m_{\rm WDM}}{1 \keV}\right)^{4/3}
\left(\frac{w_{\rm WDM}}{0.1225}\right)^{-1/3} \ .\label{eq:massrescale}
\ee
%


\section{Halo model approach}\label{sec:halomodel}

\subsection{Statistical description of the density field}

We would like to know how the nonlinear matter power spectrum changes
for WDM models. In order to do this we shall adapt the halo model
approach. For the case of CDM, the halo model would assume that all of
the mass in the Universe is in the form of dark matter haloes and so
the density field can be expressed as a sum over haloes. In WDM models
this ansatz can be no longer considered true, owing to the suppression
of halo formation on mass-scales smaller than $M_{\rm fs}$. We modify
this in the following way: we suppose that there is some fraction of
mass in collapsed dark matter haloes and that there is a corresponding
fraction in some smooth component. Hence we write the total density of
matter at a given point as:
\ba \rho(\bx) & = &
\rho_{\s}(\bx)+\sum_{i=1}^{N}M_iU(\left|x-x_i\right|,M_i) \ , \label{eq:density}\ea
where $\rho_{\s}$ is the density in smooth matter, and $U(x,M_i)\equiv
\rho(x|M)/M$ is the mass normalized isotropic density profile, and
both depend on the mass of the WDM particle -- for convenience we
suppress this label. The sum in the above extends over all $N$ haloes
and subhaloes. The mean mass density of the Universe can be written,
\be \left<\rho(\bx)\right>=\rhob=
\rhob_{\s} + \rhob_{\h} \label{eq:meandensity}\ee
where the expected total mass density in WDM haloes and the smooth
component can be written,
\be 
\bar{\rho}_{\h} = f\rhob \hspace{0.5cm} ; \hspace{0.5cm}
\bar{\rho}_{\s} = (1-f)\rhob \ , \label{eq:fracdensity}\ee
where $\rhob$ is the mean density of all matter, and $f$ the fraction
of mass in haloes. This can be written,
\be f=\frac{1}{\rhob}\int^{\infty}_{M_{\rm cut}} dM M n(M)\ , \ee
where $n(M)$ is the number density of WDM haloes per unit mass and
$M_{\rm cut}$ is the mass-scale below which there are no haloes. In
this study we will take \mbox{$M_{\rm cut}=M_{\rm fs}$}, but will
formally make the distinction throughout.

Let us now work out the clustering statistics in this model.  The two
point correlation function of the mass density can be written:
\ba 
\left<\rho(\bx)\rho(\bx+\br)\right> & = &
\left<
\left[\rho_{\s}(\bx)+\rho_{\h}(\bx)\right]
\left[\rho_{\s}(\bx+r)+\rho_{\h}(\bx+r)\right]
 \right> \nn \\
& = &  
\left<\rho_{\s}(\bx)\rho_{\s}(\bx+\br)\right>+
\left<\rho_{\s}(\bx)\rho_{\h}(\bx+\br)\right>\nn \\
& +& \left<\rho_{\h}(\bx)\rho_{\s}(\bx+\br)\right>+
\left<\rho_{\h}(\bx)\rho_{\h}(\bx+\br)\right>
\ea
Owing to the statistical homogeneity and isotropy, we have that
$\left<\rho_{\s}(\bx)\rho_{\h}(\bx+\br)\right> = \left<\rho_{\rm
h}(\bx)\rho_{\s}(\bx+\br)\right>$. The correlation function between
two random density fields X and Y can be written:
\be \xi_{\rm XY}(|\br|)\equiv\left<\delta_{\rm X}(\bx)\delta_{\rm
Y}(\bx+\br)\right>\ ,\ee
where the dimensionless over-density field $\delta$ is defined:
\be \delta_{\rm X}\equiv\frac{\rho_{\rm X}-\rhob_{\rm X}}{\rhob_{\rm X}}\ \label{eq:densitypert}
.\ee
On using the above definition, we find that
\be \xi_{\delta\delta}(r)  = 
(1-f)^2 \xi_{\ss}(r) + 2f(1-f) \xi_{\s\h}(r)  + f^2 \xi_{\hh}(r) \ .
\ee
Thus the correlation function is simply the weighted average of the
smooth-mass correlation function $\xi_{\ss}$, the mass weighted
halo-halo correlation functions $\xi_{\hh}$ and the smooth mass-halo
cross correlation function $\xi_{\h\s}$.  In order to proceed further,
we need to understand these three correlation functions.


\subsection{Halo-halo correlation function} 

Let us first consider the halo-halo correlation function, this can be
derived following \citep[][and references
therein]{CooraySheth2002}. The clustering is the sum of two terms: the
1-Halo term, which describes the clustering of points within the same
halo and the 2-Halo term, which correlates the mass distribution
between haloes. Thus we have,
\be \xi_{\hh}=\xi^{\1H}_{\hh}+\xi^{\2H}_{\hh}\ , \label{eq:xiHH}\ee
where these terms have the explicit forms \citep{CooraySheth2002}:
\ba 
\xi^{\1H}_{\hh}(r) & = & \frac{1}{\rhob_{\h}^2} 
\int_{M_{\rm cut}}^{\infty} dM M^2 n(M) \nn \\ 
& & \times \int \dx\, U(\bx|M)U(|\br-\bx|,M) \ \label{eq:xi1H} ;\\
\xi^{\2H}_{\hh}(r) & = & \frac{1}{\rhob_{\h}^2}\prod_{i=1}^{2}
\left\{ \int_{M_{\rm cut}}^{\infty} dM_i 
M_i n(M_i) \right. \nn \\ 
& & \hspace{-1.3cm}\times \left. \int \dx_i U(\br_i-\bx_i|M) \right\}
\xi^{\rm cent}_{\hh}(|\bx_1-\bx_2|)\ , \label{eq:xi2H}
\ea
where in \Eqn{eq:xi2H} $\bx_1$ and $\bx_2$ denote the position vectors
of the center of mass of halo 1 and 2, and $\br_1$ and $\br_2$ denote
position vectors of the points to be correlated, respectively. Thus
$\br_1-\bx_1$ gives the radial vector from the center of halo 1 to the
point 1, and likewise for halo 2.  Further, owing to the statistical
isotropy of the clustering $\xi(r)=\xi(|\br|)=\xi(|\br_1-\br_2|)$.

In principle the correlation function of halo centers $\xi_{\hh}^{\rm
  cent}(y|M_1,M_2)$ (with $y\equiv|\bx_1-\bx_2|$), is a complicated
scale dependent function of $M_1$, $M_2$ and $y$
\citep[e.g. see][]{Smithetal2007}.  We assume that there is a local deterministic
bias relation between haloes and dark mass, smoothed on a scale
$R$. Hence, we have
\citep{FryGaztanaga1993,MoWhite1996,Smithetal2007}:
\be \delta_{\h,R}(\bx,M)={\mathcal F}_h\!\left[\delta_R(\bx)\right]=\sum_i  
\frac{b_i(M)}{i!}\left[\delta_R(\bx)\right]^i \ \label{eq:local} ,\ee
where the $b_i(M)$ are the nonlinear bias parameters. Thus the halo
center correlation function has the form:
\ba
\xi_{\hh,R}^{\rm cent}(y) & = & b_1(M_1)b_1(M_2)\xi_R(y) \nn \\
& & \hspace{-2cm}+\frac{1}{6}\left[b_1(M_1)b_3(M_2)+b_3(M_1)b_1(M_2)\right]
\left<\delta_R(\bx)\delta^3_R(\bx+\by)\right>_{\bx} \nn \\ 
&  & \hspace{-0.5cm} 
+\frac{b_2(M_1)b_2(M_2)}{4}\left<\delta^2_R(\bx)\delta^2_R(\bx+\by)\right>_{\bx}+ 
\dots \ .\label{eq:Lin2Halo} 
\ea
where $\xi_R(r)$ is the nonlinear matter correlation function smoothed
on scale $R$. On truncating the above expression at first order in
$b_i$ and inserting into \Eqn{eq:xi2H}, one finds:
\ba \xi^{\2H}_{\hh}(r) & = & \frac{1}{\rhob_{\h}^2}\prod_{i=1}^{2}
\left\{ \int_{M_{\rm cut}}^{\infty} dM_i M_i
b_1(M_i)n(M_i)\right. \nn \\ &  & \times
\left. \int \dx_i U(\br_i-\bx_i|M) \right\}
\xi_{R}(|\bx_1-\bx_2|)\ .\ea

Before moving on, we note that in practice, in order to make accurate
predictions in the halo model on small scales, we must take into
account halo exclusion
\citep{TakadaJain2004,Smithetal2007,Smithetal2010}.  That is, the
separation of two distinct haloes can not be smaller than the sum of
their virial radii. Otherwise, they would have been identified as a
larger halo. However, when modelling WDM there is a new ``exclusion''
condition that arises, the smooth component is not really
smooth. Formally there are holes of zero density in the smooth
distribution where haloes are. Since these calculations are first
order attempts, we shall leave such detailed analysis for future
study.
%


\subsection{Smooth-smooth correlation function}

Let us next consider the smooth mass component of the model. We shall
assume that this represents all of the mass that was initially in mass
peaks of the density field that were $M<M_{\rm fs}$. In direct analogy
with the density field of dark matter haloes, we may assume that the
smooth component can be related to the total mass density field
through a deterministic local mapping. Hence,
\be \delta_{\rm s,R}(\bx)={\mathcal F}_s[\delta_R(\bx)]=
\sum_i \frac{b_{s,i}}{i!}\left[\delta_R(\bx)\right]^i\ , \label{eq:biasSmooth}\ee 
where the $b_{s,i}$ are the nonlinear bias parameters of the smooth
component.  This leads us to write an expression similar to
\Eqn{eq:Lin2Halo}, 
\be \xi_{\ss,R}(r)=b^2_{\s,1}\xi_R(r)+ {\mathcal O}(b_{s,2},b_{s,3},\dots)\label{eq:xiSS}\ .\ee


\subsection{Halo-smooth mass cross-correlation} 

The model also requires us to specify the cross-correlation between
the mass in haloes and the mass in the smooth component. This can be
written in the following fashion,
\ba 
\xi_{\h\s}(r) & = & \frac{1}{\rhob_{\h}}\int_{M_{\rm cut}}^{\infty} dM M n(M)\int \dx_1 
U(\br_1-\bx_1|M) \nn \\
& & \times b_1(M)b_{\s,1} \xi_R(\bx_1-\br_2) +{\mathcal O}(b_2,b_{s,2}\dots)
\label{eq:xiHS}\ea
where $r=|\br_1-\br_2|$.


\subsection{Power spectrum}

The correlation functions are related to the matter power spectra
through a Fourier transform
\citep{Peebles1980,Peacock1999,CooraySheth2002}:
\be
P(k) =  \int \dr\xi(r){\rm e}^{i\bk\cdot\br} \Leftrightarrow 
\xi(r) =  \int\frac{\dk}{(2\pi)^3}P(k){\rm e}^{-i\bk\cdot\br} \ . 
\ee
On applying this to Eqns~(\ref{eq:xiHH}), (\ref{eq:xiSS}), and
(\ref{eq:xiHS}), one finds
\be P_{\delta\delta}(k)=(1-f)^2P_{\ss}(k)+2(1-f)fP_{\s\h}(k)+f^2P_{\h\h}(k) \ee
where the power spectra of the smooth-smooth, smooth-halo and
halo-halo components, can be written to first order in the halo and
smooth density bias parameters as,
\ba 
P_{\s\s} & = & b_{\s,1}^2 P_{\rm NL,R} \ ; \\
P_{\s\h} & = & \frac{b_{\s,1} P_{\rm NL,R}}{\rhob_{\h}}
\int_{M_{\rm cut}}^{\infty} dM M b_1(M) n(M) \tilde{U}(k|M)\ ; \\ 
P_{\h\h} & = & \frac{P_{\NL,R}}{\rhob_{\h}^2}\left[
\int_{M_{\rm cut}}^{\infty} dM M b_1(M) n(M) \tilde{U}(k|M)\right]^2  \nn \\ 
& & + \frac{1}{\rhob_{\h}^2}\int_{M_{\rm cut}}^{\infty}dM n(M) M^2 \tilde{U}^2(k|M)\ , \ea 
where $P_{\rm NL,R}\equiv P_{\rm NL,R}(k)$ is the nonlinear mass power
spectrum smoothed on a scale $R$, and $ \tilde{U}(k|M)$ is the Fourier
transformed mass normalized density profile, $U(x,M_i)$.


\subsection{Bias of the smooth component}\label{ssec:smoothbias}

Absent so far is any explanation as to how one might infer the bias
factors of the smooth component $b_{\s,i}$. We shall now use a mass
conservation argument to derive a relation between the bias of the
smooth mass and the bias of the haloes that involves no new free
parameters. That is, if we fully understand halo biasing, then we may
immediately obtain a full understanding of the density distribution of
the smooth density component.

Consider again the density at a given point, this may be written as in
\Eqn{eq:density}, i.e. $\rho(\bx)=\rho_{\s}(\bx)+\rho_{\h}(\bx)$.  On
using Eqns~(\ref{eq:meandensity}), (\ref{eq:fracdensity}) and
(\ref{eq:densitypert}), then we may write this in terms of the
overdensities in the smooth and clumped matter as
\be 
\delta(\bx) = (1-f)\delta_{s}(\bx)+f\delta_{\h}(\bx)\label{eq:totalden}\ .
\ee
The mass density associated with matter in haloes can be written as an
integral over all haloes above the mass-scale $M_{\rm cut}$ in the
following manner,
\ba 
\rho_{\h}(\bx) & = & \rhob_{\h}\left[1+\delta_{\h}(\bx)\right] \nn \\
& = & \int_{M_{\cut}}^{\infty} dM M n(M)
\left[1+\delta_{\h}(\bx|M)\right] \ .
\ea
If we assume the local model for halo biasing of \Eqn{eq:local}, then
the density field of haloes takes the form:
\be 
\delta_{\h}(\bx)  =  \frac{1}{\rhob_{\h}}
\int_{M_{\cut}}^{\infty} dM M n(M)\sum_i\frac{b_i(M)}{i!}\left[\delta_R(\bx)\right]^i
\label{eq:biasM}\ee
where $\rhob_{\h}\equiv f \rhob$. Next, let us define the effective
mass weighted halo bias parameters as,
\be 
b^{\rm eff}_i=\frac{\int_{M_{\cut}} dM M n(M) b_i(M)}{\int_{M_{\cut}} dM M n(M)}\ .
\ee
Then we may re-write \Eqn{eq:biasM} more simply as 
\be 
\delta_{\h}(\bx)=\sum_i \frac{b^{\rm eff}_i}{i!} \left[\delta_R(\bx)\right]^i\ .
\ee
Substituting the above equation and the local bias model from
\Eqn{eq:biasSmooth} into \Eqn{eq:totalden}, we then find
\ba
\delta(\bx) & = & (1-f)\sum_i \frac{b_{s,i}}{i!} \left[\delta_R(\bx)\right]^i+ 
f \sum_i \frac{b^{\rm eff}_i}{i!} \left[\delta_R(\bx)\right]^i. \nn \\
& = & \sum_i \left[ (1-f)b_{s,i} + f b^{\rm eff}_i\right]
 \frac{\left[\delta_R(\bx)\right]^i}{i!}\ .
\ea
On inspecting the above equation, two conditions may be immediately
noticed. Firstly, in order for the left and right sides of the
equation to balance, we must have that 
\be 1 = (1-f)b_{\s,1}+fb_1^{\rm eff}\ .\ee 
This leads us to the result that the bias of the smooth component must
be given by,
\be b_{\s,1}=\frac{1-fb_1^{\rm eff}}{1-f}\label{eq:bsmooth} .\ee 
Secondly, if the smooth component is indeed linearly biased with
respect to the total matter overdensity, then we must also have that
the nonlinear bias coefficients vanish when integrated over all
haloes:
\be \int_{M_{\cut}} dM M n(M)b_i(M)=0 \ (i\ne1)\ .\ee
On the other hand, supposing that the smooth component is not simply
linearly biased, but is nonlinearly biased {\em a la} the local model,
then the above condition is relaxed to the condition:
\be 
(1-f)b_{\s,i}+fb^{\rm eff}_{i}=0 
\ \ (i\ne1)\ .
\ee
This leads us to the condition,
\be b_{\s,i}=-\frac{fb_{\h,i}^{\rm eff}}{1-f}\ \ (i\ne1)\ .\ee
Hence, if the haloes and the smooth mass component are locally biased,
then all of the nonlinear bias parameters for the smooth component can
be derived from knowledge of the halo bias parameters. In
\S\ref{ssec:halobias} we will show how the bias of the smooth
component varies for different mass-scales of the WDM particle.


\begin{figure*}
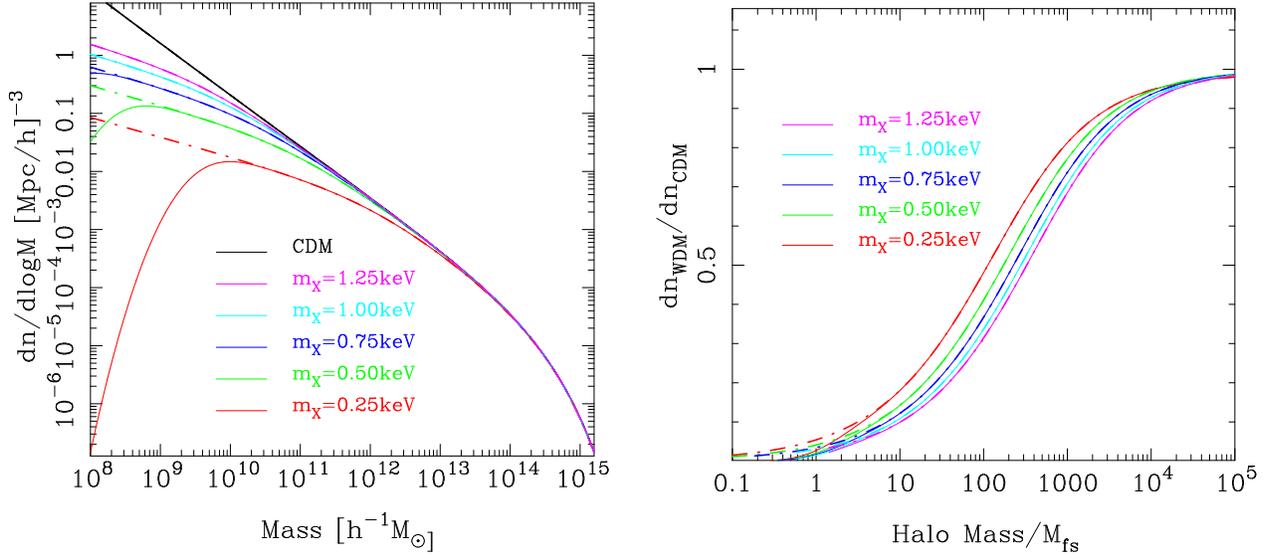

\centering{ 
\includegraphics[width=8.0cm,clip=]{FIGS/MassFunWDM.1.ps}\hspace{0.5cm}
\includegraphics[width=8.0cm,clip=]{FIGS/MassFunWDM.Scale.1.ps}}
\caption{\small{{\em Left panel}: Halo mass function as a function of
halo mass. Lines show the theoretical predictions from the ST model of
\Eqn{eq:STMF}. The black solid line denotes the result for CDM; the
dot-dash lines denote results for WDM for our standard set of particle
masses ($m_{\WDM}\equiv m_{X}=\{0.25,\,0.5\,0.75\,1.0,\,1.25\}\,{\rm
keV}$) with no cut-off mass applied; the solid colored lines denote
the same, but with a cut-off mass-scale as given in \Eqn{eq:mfcutoff},
and here we use $\sigma_{\log M}=0.5$. {\em Right panel}: Fractional
difference between the WDM and CDM mass functions, as a function of
halo mass scaled in units of the WDM free-streaming mass-scale $M_{\rm
fs}$. Line styles are the same as in left panel.}\label{fig:mf}}
\end{figure*}


\begin{figure*}
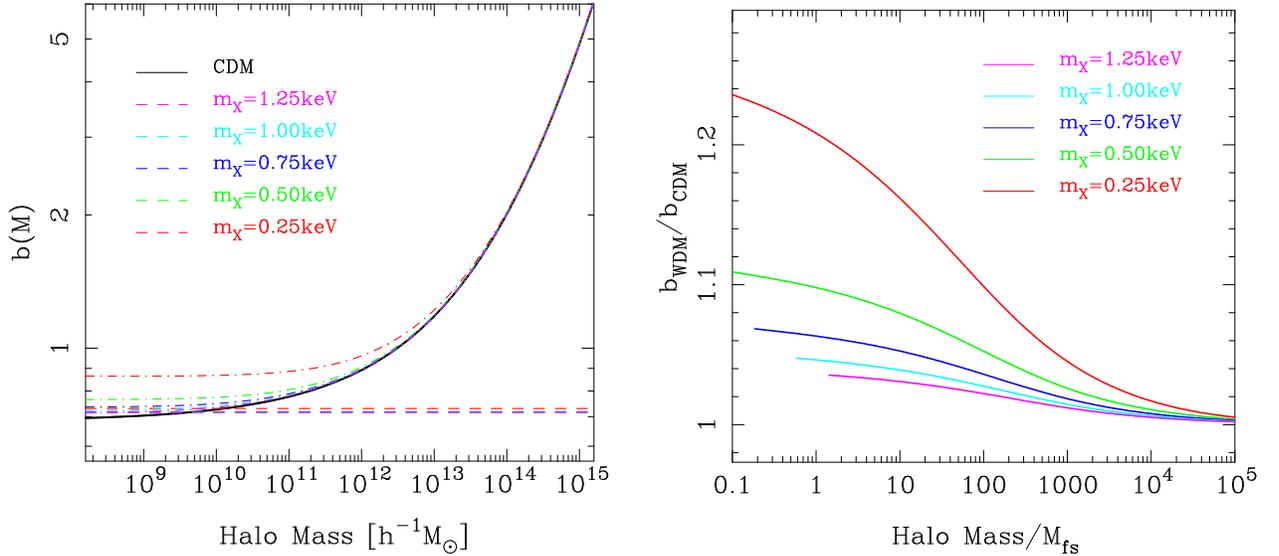

\centering{
  \includegraphics[width=8.0cm,clip=]{FIGS/biasWDM.ps}\hspace{0.5cm}
  \includegraphics[width=8.0cm,clip=]{FIGS/biasWDM.Scale.ps}}
\caption{\small{{\em Left panel}: Halo bias as a function of halo mass
for various dark matter models. The solid black line denotes the
results for CDM; the dot-dashed line denotes the halo bias for our
standard set of WDM particle masses
($m_{\WDM}\equiv m_{X}=\{0.25,\,0.5,\,0.75,\,1.0,\,1.25\}\,{\rm keV}$).} The dashed
lines denote the bias of the smooth component of mass as given by
\Eqn{eq:bsmooth}. {\em Right panel}: fractional difference between the
bias in WDM and CDM models as a function of halo mass-scaled in units
of the free-streaming mass-scale $M_{\rm fs}$. \label{fig:bias}}
\end{figure*}


\section{Halo Model ingredients in WDM models}\label{sec:ingredients}

In this section we detail the halo model ingredients and show how they
change in the presence of our benchmark set of WDM particle masses.


\subsection{Halo mass function}\label{sec:ingredientsI}

In order to predict the abundance of haloes in the WDM cosmological
model, we shall adopt the Press-Schechter (PS)
\citep{PressSchechter1974} approach, with some small modification.
For mass-scales $M>M_{\rm fs}$, we shall assume that the standard
formalism holds. In this case we may calculate the abundance of
objects through the usual recipe, only being sure to use the
appropriate linear theory power spectrum for WDM. Hence,
\be \frac{dn}{d\log M}=-\frac{1}{2}\frac{\rhob}{M} f(\nu)
\frac{d\log\sigma^2}{d \log M}\ ;
\hspace{1cm} \nu\equiv \frac{\delta_c(z)}{\sigma(M)}\ ,\ee
where $dn=n(M)dM$ is the abundance of WDM haloes of mass $M$ in the
interval $dM$, $\delta_c(z)=1.686/D(z)$ is the linear theory collapse
barrier for the spherical collapse model, and $D(z)$ is the linear
theory growth function. Note that we take the evolution of
$\delta_c(z)$ and $D(z)$ to be unchanged from their forms in the
$\Lambda$CDM model. Our reasoning is that, if one follows the standard
derivations for $D(z)$ and $\delta_c(z)$, then, provided the WDM
particles decouple early and are non-relativistic today, these
functions are solely determined from the dynamics of the expansion
rate of the Universe. This assumption will of course break down at
very early times, when the primordial thermal velocities are more
significant and when the particles are relativistic and so growth
suppressed.

For the function $f(\nu)$ we adopt the model
of \citet{ShethTormen1999}, and this has the form:
\be
f_{\rm ST}(\nu) =  
      A(p) \sqrt{\frac{2}{\pi}} \sqrt{q}\nu \left[1+(\sqrt{q}\nu)^{-2p}\right]
\exp\left[-\frac{q\nu^2}{2}\right] \label{eq:STMF}\ .
\ee
For the ST mass function the amplitude parameter $A$ is determined
from the constraint $\int d\log \nu f(\nu) =1$, which leads to
$A^{-1}(p)=\left[1+2^{-p}\Gamma\left[0.5-p\right]/\Gamma\left[0.5\right]\right]$.
The parameters are: $\{A=0.3222, p=0.3, q=0.707\}$. The variance on
mass-scale $M$ and its derivative with respect to the mass scale can
be computed from the following integrals:
\ba 
\sigma^2(M) & = & \int \frac{\dk}{(2\pi)^3} P_{\rm Lin}(k) W^2\ ; \label{eq:sig} \\
\frac{d\log\sigma^2(R)}{d\log M} & = & \frac{2}{3\sigma^2(R)}\int
\frac{\dk}{(2\pi)^3} P_{\rm Lin}(k) W W' \ . \label{eq:dsig}
\ea
where the filter functions are $W\equiv W(kR)$ and $W' \equiv
dW(kR)/d\log (kR)$.  For a 3D real-space spherical top-hat filter,
transformed into Fourier space, we have:
\ba 
W_{\rm TH}(y) 
& = & \frac{3}{y^3}\left[\sin y - y \cos y\right] \ ; \\
W_{\rm TH}'(y)
& = & \frac{3}{y^3}\left[(y^2-3)\sin y + 3y \cos y\right] \ .
\ea

For mass-scales $M<M_{\rm fs}$, free-streaming erases all peaks in the
initial density field, and hence peak theory should tell us that there
are no haloes below this mass scale. A blind application of the PS
approach leads one to predict, suppressed, but significant numbers of
haloes below the cut-off mass. 

In order to demonstrate this point explicitly, let us consider the
following toy-model calculation. Suppose that the matter power
spectrum is a simple power-law $P(k)\propto k^{n}$, and let us suppose
that if the dark matter particle is warm then the effect of the
damping on the fluctuations due to free-streaming can be represented
by a simple exponential damping. Hence, $P_{\rm
  WDM}(k)=A_s(k/k_c)^{n}\exp[-k/k_c]$, where $k_c$ is a characteristic
damping scale; perhaps $k_c=2\pi/\lambda_{\rm fs}$. Let us now
consider the behaviour of \Eqns{eq:sig}{eq:dsig} in the limit that
$R\rightarrow0$.  For the variance term we have:
\ba
\lim_{R\rightarrow0} \sigma^2(R) 
& = & \int \frac{\dk}{(2\pi)^3} P_{\WDM}(k) \lim_{R\rightarrow0} W^2(kR)\nn \\
& = & \frac{4\pi A_s k_c^3}{(2\pi)^3} \Gamma[n+3]
\ea
where we have used the fact that $\lim_{R\rightarrow0} W(kR)=1$. For
the derivative of the variance we have:
\ba 
\lim_{R\rightarrow0} \frac{d\log\sigma^2(R)}{d\log M} & = & 
\left[\lim_{R\rightarrow0} \frac{2}{3\sigma^2(R)}\right]
 \int\frac{\dk}{(2\pi)^3} P_{\WDM}(k) \nn \\
& & \times \left[\lim_{R\rightarrow0}W(kR)\right] 
\left[\lim_{R\rightarrow0}W'(kR) \right]  \nn \\
& = & -1\ ,
\ea
where we used the fact that $\lim_{R\rightarrow0}W'(kR)=-3/2$. Thus as
$M\rightarrow0$ (or equivalently $R\rightarrow0$), the mass function
of dark matter haloes becomes:
\ba 
\lim_{M\rightarrow0}\frac{dn}{d\log M} & = & \lim_{M\rightarrow0}
\left[\frac{1}{2}\frac{\rhob}{M} \right] A(p) \sqrt{\frac{2}{\pi}} \sqrt{q} \nu_0 \nn \\
&  & \hspace{-1.5cm}\times  \left[1+(\sqrt{q}\nu_0)^{-2p}\right] 
\exp\left[-\frac{q\nu_0^2}{2}\right] \rightarrow \infty\ ,
\ea
where in the above we have defined
\be
\nu_0\equiv \frac{\delta_c(z_c)}{\sqrt{B\Gamma[n+3]}}\ ; \ \ 
\ B\equiv\frac{4\pi A_sk_c^3}{(2\pi)^3}\ .
\ee
Hence, we see that the number density of haloes per logarithmic
interval diverges, $\lim_{M\rightarrow0}M n(M)\rightarrow \infty$, but
that the multiplicity function of haloes asymptotes to a constant
value, $\lim_{M\rightarrow0}M^2 n(M)\rightarrow \rm const$.

We show this behaviour more schematically in the left panel of
\Fig{fig:mf} (dot-dashed curves), where we compare CDM and WDM mass
functions. For large halo masses $M>10^{13}\Msol$, the models are
virtually indistinguishable, even for the case of a WDM particle with
mass $m_{\WDM}=0.25\keV$. However, for smaller halo masses, whilst
there is a significant suppression, they are still highly
abundant.

Recent results from numerical simulations show that there is a strong
suppression in the numbers of haloes with $M<M_{\rm fs}$
\citep{Zavalaetal2009,PolisenskyRicotti2010}. Also, there is some
debate as to whether haloes below the cut-off can form at all
\citep{Bodeetal2001,WangWhite2007}. We shall take the view that PS is
correct for $M>M_{\rm fs}$, but that below this mass one finds
virtually no haloes. We model this transition in an {\em ad-hoc}
manner using an error function as a smoothed step. Thus the mass
function in WDM becomes:
\be \frac{d\tilde{n}_{\WDM}}{d\log M}= \frac{1}{2}\left\{1+{\rm erf}
\left[ \frac{ \log_{10}(M/M_{\rm fs})}{\sigma_{\log M}} \right]\right\}  
\frac{dn_{\WDM}}{d\log M}\ ,\label{eq:mfcutoff}\ee
where $\sigma_{\log M}$ controls the logarithmic width of the step,
and $M_{\rm fs}$ controls the location of the step. In this work, for
simplicity we shall take $\sigma_{\log M}\rightarrow0$, and so the
function acts like a standard Heaviside function at $M_{\rm fs}$.  We
show the results of this modification in \Fig{fig:mf} (solid
lines). For this example we have chosen $\sigma_{\log M}=0.5$.

The exact amount of suppression in the WDM mass functions, relative to
CDM, can be better quantified through inspecting the right panel of
\Fig{fig:mf}. Here we show the fractional difference between the WDM
and CDM models versus halo mass-scaled in units of $M_{\rm fs}$.  The
plot shows that there is roughly a $\sim50\%$ suppression in the
abundance of haloes with \mbox{$M\lesssim100-300 \, M_{\rm fs}$} in
all of the WDM models considered.


\subsection{Halo bias}\label{ssec:halobias}

The halo model also requires us to specify how the centers of dark
matter haloes of different masses cluster with respect to each other.
As described by \Eqn{eq:Lin2Halo}, for the Gaussian model, and at
first order in the dark matter density, halo and matter density
perturbations can be related through a scale-independent bias factor
$b(M)$.  Following \citep{MoWhite1996,ShethTormen1999}, an application
of the peak-background split approximation enables one to calculate
$b(M)$ from a given mass function. For the ST mass function the
Gaussian bias has the form:
\be b_{\rm ST}(\nu)= 1 + \frac{q\nu^2
  -1}{\delta_c(z)}+\frac{2p}{\delta_c(z)\left[1+(q\nu^2)^p\right]} \ ,
\label{eq:halobias}\ee
where the parameters $\{p,q\}$ are as in \Eqn{eq:STMF}.

For the case of WDM, we shall assume that for haloes with $M>M_{\rm
fs}$ the bias is well described by the same peak-background split
model. Hence in model space: $b_{\WDM}(M)\rightarrow b_{\CDM}(M)$.

In the left panel of \Fig{fig:bias} we show how the bias of dark
matter haloes in both CDM and WDM models depends on the halo mass. The
figure shows the results for the set of WDM particle masses:
$m_{\WDM}=\{0.25,\,0.5,\,0.75,\,1.0,\,1.25\}\,{\rm keV}$.  As for the
mass function, for high mass haloes with $M>10^{13}\Msol$, the bias is
virtually indistinguishable from that obtained for CDM. For the case
of lower masses, we see that the bias has a clear dependence on the
mass of the WDM particle. For lower mass particles the bias of low
mass haloes is significantly boosted with respect to CDM.

The right panel of \Fig{fig:bias} shows this last point in more
detail. Here we plot the ratio of the bias in the WDM model with
respect to that of the CDM model, and we scale the halo mass in terms
of the free-streaming mass-scale $M_{\rm fs}$. We now see that there
is roughly a $10\%\,(2\%)$ boost in the bias for haloes with
$M\lesssim 100\,M_{\rm fs}$ and with a WDM particle mass
$m_{\WDM}=0.25\, (1.25) \keV$.

The left panel of \Fig{fig:bias} also shows how the bias of the smooth
mass distribution varies with respect to $m_{\WDM}$. This was
calculated using \Eqn{eq:bsmooth}, and assuming the ST mass function
and bias parameter from above. We find that the smooth component is
anti-biased with respect to the total matter, with \mbox{$b_s\in
\left[0.730,\,0.718,\,0.716,\,0.715,\,0.715\right]$} for the cases
\mbox{$m_{\WDM}=\{0.25,\dots,1.25\}\,{\rm keV}$}, respectively.  Note
that as the particle mass increases the bias of the smooth component
saturates. This arises due to the fact that, for $M/M_*\ll 1$,
$\nu\rightarrow0$ and hence from \Eqn{eq:halobias},
\ba 
\lim_{\nu\rightarrow 0} b_{\rm ST}(\nu) & = & 1+\left[2p-1\right]/\delta_c(z)\nn \\
& \approx & 1-D(z)\frac{0.4}{1.686} \approx 0.7 \ (z=0)\ .
\ea


\begin{figure*}
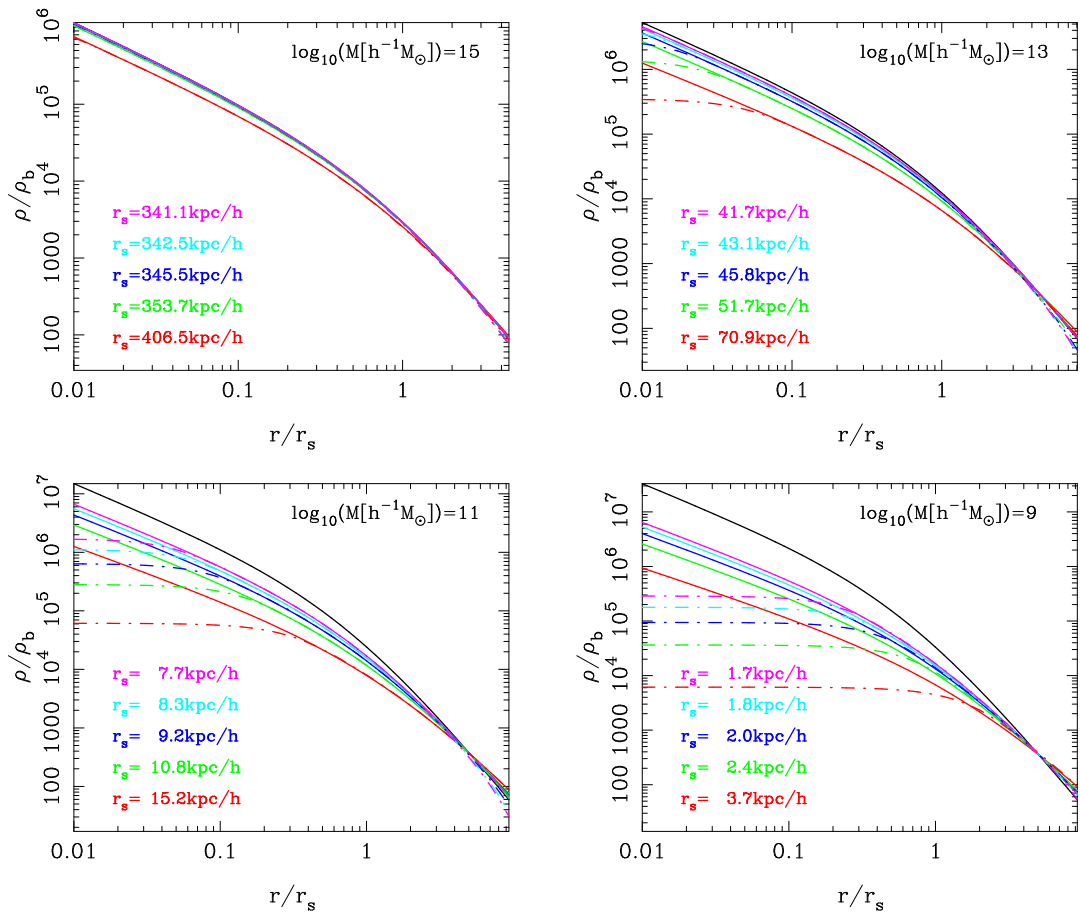

\centering{
  \includegraphics[width=6.7cm,clip=]{FIGS/ProfileWDM.15.ps}\hspace{0.75cm}
  \includegraphics[width=6.7cm,clip=]{FIGS/ProfileWDM.13.ps}}\vspace{0.3cm}
\centering{
  \includegraphics[width=6.7cm,clip=]{FIGS/ProfileWDM.11.ps}\hspace{0.75cm}
  \includegraphics[width=6.7cm,clip=]{FIGS/ProfileWDM.9.ps}}
\caption{\small{Density profiles of CDM and WDM haloes as a function
of radius scaled in units of the characteristic scale radius
$r_s$. Top left, top right, bottom left and bottom right panels show
the results for haloes with masses: $M \in
\{10^{15},\,10^{13},\,10^{11},\,10^{9}\}\Msol$. In all panels: the
black line denotes the CDM profile; the (red, green, blue, cyan,
magenta) lines denote the WDM profiles for masses \mbox{$m_{\WDM}
\in\{0.25,\,0.5,\,0.75,\,1.0,\,1.25\}\,{\rm keV}$}, respectively. The
solid lines denote the case where there are no thermal relic
velocities; the dot-dash lines denote the case where profiles have
been smoothed with a Gaussian filter of the scale of the mean relic
velocity speed $l_{v_{\rm r}}\sim\overline{v}/H_0\sim 3.15v_0/H_0$. In
each panel we also indicate the scale radius that the halo would have
for the various WDM particle masses considered, where from top to
bottom we show this for decreasing $m_{\WDM}$. }\label{fig:profile}}
\end{figure*}


\subsection{Halo density profiles}

The density profiles of CDM haloes have been studied in great detail
over the last two decades.  A reasonably good approximation for the
spherically averaged density profile in simulations is the NFW model
\cite{Navarroetal1997}:
\be \rho^{\rm NFW}_{\CDM}(r|M)=\rhob\,\Delta_{\rm c}(M) \left[
  y\left(1+y\right)^2 \right]^{-1} \ ;\ \ y\equiv r/r_{\rm s} ,\ee
where the two parameters are the scale radius $r_{\rm s}$ and the
characteristic density $\Delta_c$. If the halo mass is $M_{\rm
  vir}=4\pi r_{\rm vir}^3200\rhob/3$, then owing to mass conservation
there is only one free parameter, the concentration parameter
$c(M)\equiv r_{\rm vir}/r_{\rm s}$. Hence,
\be \Delta_{\rm c}(M)=(200/3)
c^3\left[\log\left(1+c\right)-c/(1+c)\right]^{-1}\ .\ee
The concentration parameter can be obtained from the original model of
NFW, but we shall instead use the model of
\citet{Bullocketal2001}. Note that for our halo model calculations we
shall correct $c(M)$ for the fact that the definitions of the virial
mass in Bullock et al. and Sheth \& Tormen are different
\citep[][]{SmithWatts2005}.

In the case of the WDM model, there are two major effects which modify
the density profiles when compared to CDM. Firstly, if one adopts
either the NFW~\citep{Navarroetal1997} or
Bullock~et~al.~\citep{Bullocketal2001} approach for calculating the
concentration parameters, then one finds, owing to the suppression of
small-scale power, that the collapse redshifts of the haloes are
significantly affected. Thus a given halo of mass $M$ in the WDM
model, will be expected to collapse at a later time than in the CDM
model. Since halo core density is related to the density of the
Universe at the collapse time, then one expects that core densities in
WDM models will be suppressed, and haloes on average to be less
concentrated.

We explicitly demonstrate this point in \Fig{fig:profile}. The top
left, top right, bottom left, and bottom right panels show density
profiles for haloes of mass
\mbox{$M\in\left[10^{15}\,\,10^{13},\,10^{11},\,10^{9}\right]\Msol$},
respectively. The CDM case is shown as the black solid line, and the
case where the mass of the dark matter particle is decreased, is shown
as the series of curves with decreasing amplitude. Clearly, smaller
mass haloes are more significantly affected by the absence of
small-scale power. However, we do note that higher mass haloes are not
immune to this, there being signs of density suppression in haloes with
$M\sim10^{13}\Msol$ for $m_{\WDM}\sim1 \keV$.

Secondly, it has been argued, owing to the preservation of Liouville's
theorem and the existence of a relic thermal velocity distribution for
the dark matter particles, that there is a fundamental limit on the
size of core densities in phase space
\citep{TremaineGunn1979,Colinetal2008}. If this is correct, then cuspy
density profiles are prohibited in the WDM model. In order to mimic
the effect of the phase space constraint on the profiles, we take a
qualitative approach and simply convolve the NFW profile (obtained
from computing collapse times with the WDM power spectrum) with a
Gaussian filter function of radius the mean thermal velocity scale of
the WDM particles.

This can be calculated as follows: Let us suppose that the WDM
particles are leptons and so obey the Fermi-Dirac (FD)
statistics. Also, we assume that the gas decoupled from the rest of
matter whilst still relativistic, i.e. the energy of a given particle
is $E\sim c p$. Thus its phase space distribution remains unchanged,
except cooling adiabatically through the expansion of the
Universe. Hence, the FD distribution function can be written:
\be F(v) \propto \left(1+\exp[v/v_0]\right)^{-1}\ .\ee
Fitting to the results from the evolution of Einstein-Boltzmann codes
gives the characteristic velocity (temperature) to be
\citep{Bodeetal2001}:
\ba
v_0(z)& \approx & 0.012 (1+z)\left[\frac{\Omega_{\WDM}}{0.3}\right]^{1/3} 
\left[\frac{h}{0.65}\right]^{2/3}
\nn \\
& & 
\times \left[\frac{g_{\WDM}}{1.5}\right]^{-1/3} 
\left[\frac{m_{\WDM}}{1\keV}\right]^{-4/3} {\kms}\ ,
\ea
where $g_{\WDM}$ is the number of degrees of freedom per particle, and
we take this to be similar to that of a light neutrino species, hence
$g_{\WDM}\sim1.5$ \citep{Bodeetal2001}.  Next, in order to calculate
statistics, we also require the density of momentum states, for a
quantum gas this is: $N(p)\propto d^3p/h^3\propto d^3v\propto
N(v)$. Hence, the mean and rms velocities can now be calculated:
\ba 
\overline{v_{\rm r}} & = & A\int_0^{\infty} dv v N(v) F(v) =  \frac{7\pi^4}{180\zeta(3)} v_0
\label{eq:mean} \
; \\
\overline{v^2_{\rm r}} & = & A \int_0^{\infty} dv v^2 N(v) F(v) = 
\frac{15 \zeta(5)}{\zeta(3)}  v_0^2 \ .
\label{eq:rms}
\ea
In order to obtain the last equalities in both \Eqns{eq:mean}{eq:rms},
we have used the normalization condition,
\be 
\frac{1}{A} =  \int_0^{\infty} dv v^2 \left(1+{\rm e}^{v/v_0}\right)^{-1} = \frac{3}{2}\zeta(3)v_0^3\ .
\ee
$\zeta(3)\approx1.202$ and $\zeta(5)\approx1.037$ are Riemann Zeta
functions. Evaluation of \Eqns{eq:mean}{eq:rms} gives:
\mbox{$\overline{v_{\rm r}}=3.15 v_0$} and $\sqrt{\overline{v_{\rm
r}^2}}=3.60v_0$.  Table~\ref{tab:FDstats} presents the velocity
statistics for our fiducial set of WDM particle masses.
\begin{table}
\caption{Variation of the present day ($z=0$) relic velocity
statistics with WDM particle mass. Columns are: particle mass,
characteristic velocity of FD gas, mean velocity, rms velocity and the
smoothing scale for profiles.
\label{tab:FDstats}}
\begin{ruledtabular}
\begin{tabular}{ccccc}
$m_{\WDM}$ & $v_0$ &  $\overline{v_{\rm r}}$ & $\sqrt{\overline{v_{\rm r}^2}}$ & $l_{v_{\rm r}}$ \\
 $[\keV]$  & $[\kms]$ & $[\kms]$ & $[\kms]$ & $[\kpc]$ \\
\hline
0.25 & 0.065 & 0.204 & 0.234 & 2.04\\
0.50 & 0.026 & 0.082 & 0.094 & 0.82\\
0.75 & 0.015 & 0.047 & 0.054 & 0.47\\
1.00 & 0.010 & 0.032 & 0.036 & 0.32\\
1.25 & 0.008 & 0.025 & 0.029 & 0.25
\end{tabular}
\end{ruledtabular}
\end{table}

Finally, the density profile smoothed on the scale of the mean relic
velocity field, is given by
\be {\tilde\rho}_{\WDM}^{\rm NFW}(r|M)=
\int \frac{\dk}{(2\pi)^3}M U^{\rm NFW}_{\CDM}(k|M) W_{\rm G}(k|l_{v_{\rm r}}) 
j_0(kr) \ , \label{eq:FourierProfile}\ee
where $l_{v_{\rm r}}\approx \overline{v_{\rm r}}(z)/H_0$, $W_{\rm G}(k|l_{v_{\rm r}})\equiv 
\exp \left[-(k l_{v_{\rm r}})^2/2\right]$ and where $U^{\rm
NFW}_{\CDM}(r|M)\equiv \rho^{\rm NFW}_{\CDM}(r|M)/M$ is the mass
normalized profile. We truncate the CDM profile at the virial radius
and for the NFW model the Fourier transform can be written as
\citep{CooraySheth2002}:
\ba
f(c) U_{\CDM}^{\rm NFW}(k|M) & = & -\frac{\sin(kcr_{\rm s})}{kr_{\rm s}(1+c)}\nn \\
& & \hspace{-2.4cm}
\frac{}{}+\cos\left[kr_{\rm s}(1+c)\right]\left\{C_i\left[kr_{\rm s}(1+c)\right]
-C_i[kr_{\rm s}]\right\}\nn \\
& & \hspace{-2.4cm} 
\frac{}{}+\sin\left[kr_{\rm s}(1+c)\right]\left\{S_i\left[kr_{\rm s}(1+c)\right]
-S_i[kr_{\rm s}]\right\}, 
\label{eq:FourierProfileNFW}\ea
where $f(c)\equiv{[\log\left(1+c\right)-c/(1+c)]}$ and where $S_i$ and
$C_i$ are the standard sine and cosine integrals. 

In \Fig{fig:profile} we show the impact of relic velocities on the
density profiles. If the relic velocities are present and act to
simply smooth the density field, then, by {\em fiat}, we see that halo
cusps are suppressed. Importantly, the figure shows, for the range of
WDM particle masses considered, that only haloes with
$M\lesssim10^{11}$ show any noticeable effects due to the relic
velocities on scales $r\gtrsim 0.1 r_s$.


\section{Results}\label{sec:results}


\begin{figure}
\centering{
  \includegraphics[width=8.0cm,clip=]{FIGS/PowHaloModelWDM.AllComponents.1.ps}}
\caption{\small{Comparison of the nonlinear matter power spectrum in
    the halo model for a particular WDM model \mbox{($m_{\WDM}\equiv
      m_{x}=0.25 \keV$)} and the CDM model, as a function of
    wavenumber. The solid black and red lines denote the total CDM and
    total WDM power spectra in the halo model. The dashed, dot-dash,
    dotted and triple dot-dashed lines denote, $P_{\s\s}$, $P_{\s\h}$,
    $P_{\1H}$, and $P_{\2H}$, respectively. The solid blue and red
    lines correspond to results computed with and without the relic
    velocities of the WDM particles -- note that these lines are
    indistinguishable in the plot.}\label{fig:powall}}
\end{figure}


\begin{figure*}
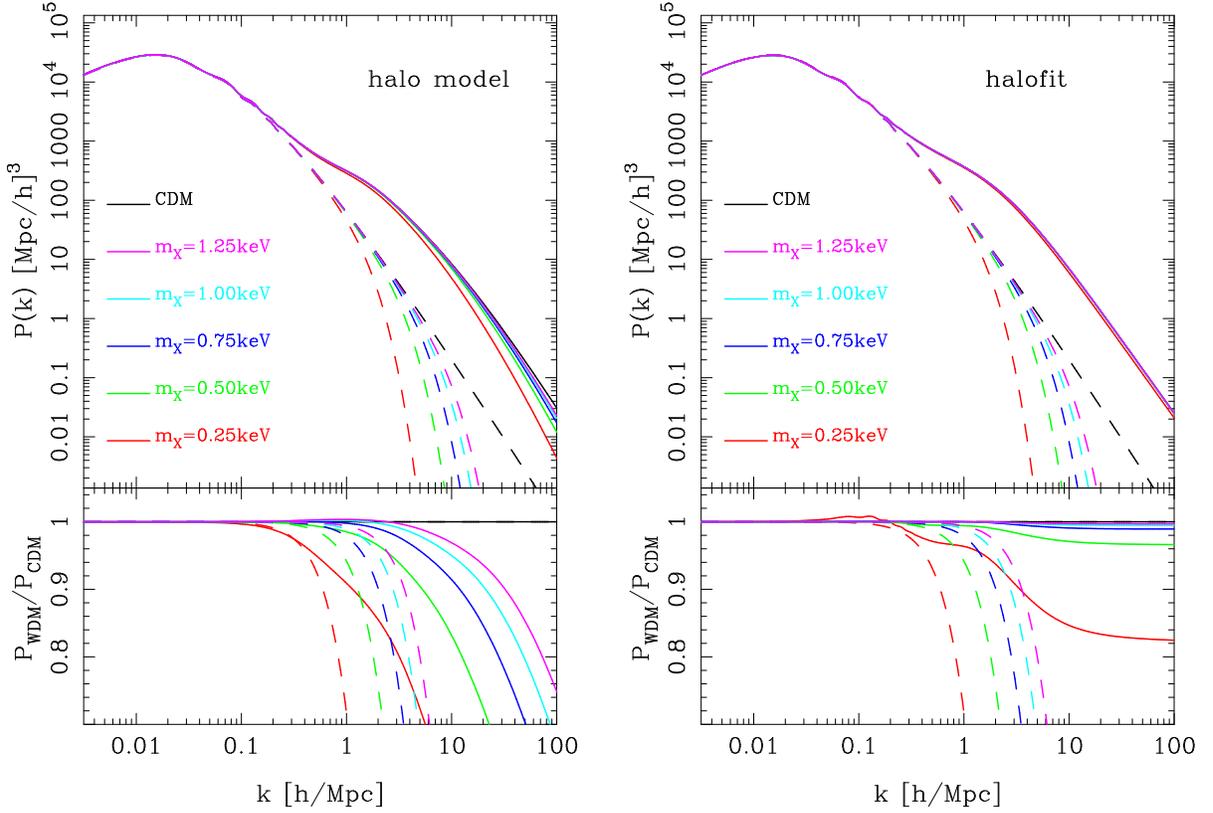

\centering{
  \includegraphics[width=7.6cm,clip=]{FIGS/PowHaloModelWDM.HM.ps}\hspace{0.5cm}
  \includegraphics[width=7.6cm,clip=]{FIGS/PowHaloModelWDM.halofit.ps}}
\caption{\small{Comparison of the halo model predictions for the WDM
    and CDM matter power spectra as a function of wavenumber. {\em
    Left and right panels}: results as obtained from the Halo Model
    and {\tt halofit}, respectively. {\em Top panels}: absolute
    power. Solid and dashed lines denote the nonlinear and linear
    theory predictions, respectively. Black lines denote CDM; and
    colors \{red, green, blue, cyan, magenta\} denote WDM particle
    masses \mbox{$m_{\WDM}\equiv
    m_{X}\in\{0.25,\,0.5,\,0.75,\,1.0,\,1.25\}\,{\rm keV}$}. {\em
    Bottom panels:} ratio of WDM and CDM power spectra. Line styles
    unchanged.  }\label{fig:powerHM}}
\end{figure*}


\subsection{Mass power spectra}

In \Fig{fig:powall} we compare the WDM and CDM nonlinear matter power
spectra obtained from the halo model calculations.  The figure shows
the contribution of each of the halo model terms to the WDM
spectrum. For the purposes of clarity we show this for one case only,
a WDM particle with $m_{\WDM}=0.25\,\keV$. This is instructive, as it
shows that on large scales the power is totally determined by the
weighted contributions $P_{\s\s}$, $P_{\s\h}$ and $P_{\2H}$. However,
on small scales the power is entirely dominated by the haloes that
exist above the free-streaming mass-scale. The integrand of the 1-Halo
term is $n(M)M^2|U(k|M)|^2$. Since there are virtually no differences
between the mass function in WDM and CDM for haloes above $M\sim
10^{13}\Msol$, the apparent suppression in power comes entirely from
the suppression in the abundance of small haloes and the change in the
profiles.

In \Fig{fig:powall} we also show the importance of the relic
velocities on the matter power spectrum. As can clearly be seen, at
present times the halo model power spectra with and without the relic
velocity effects are indistinguishable for scales $k<100\kMpc$.  On
these scales, the modifications to the power spectrum due to
uncertainties in the baryonic physics are significantly more
important. We therefore assume that the present day effects of relic
velocities will be unimportant for constraining the WDM particle from
large-scale structure probes.

In the left panel of \Fig{fig:powerHM}, we show the dependence of the
nonlinear matter power spectra on the WDM particle mass. We see that,
whilst the linear theory power spectra are significantly damped, the
nonlinear halo model predictions show that a large amount of small
scale power is generated. As noted above this arises due to the
collapse of haloes above the free-streaming mass-scale. The bottom
panel of \Fig{fig:powerHM} allows us to quantify the differences in
greater detail. Here we show the ratio of the linear WDM and CDM
spectra, and also the ratio of the nonlinear WDM and CDM spectra. From
this we understand that the relative difference in the nonlinear model
is significantly less than the difference between the linear theory
predictions. For our cosmological model, the effect of going from the
CDM model to WDM model with particle mass $m_{\WDM}\gtrsim 1\keV$, is
to cause a $\sim10\%$ suppression of power at $k\gtrsim30\kMpc$.

In the right panel of \Fig{fig:powerHM}, we show the nonlinear
predictions for the WDM power spectra from the {\tt halofit} code of
\citep{Smithetal2003}. This semi-empirical method was designed to
match CDM power spectra. It is not clear whether this method can be
accurately extrapolated to model nonlinear WDM spectra. Comparing the
left and right panels of the figure, we see that the halo model
predicts a stronger suppression of nonlinear power than {\tt halofit},
and also a greater difference between the predictions for each WDM
particle.  It will require hi-resolution $N$-body simulations, to
disentangle which of these two models provides the better description
for the power spectrum in WDM models.

At this juncture, one might question the validity of the halo model
predictions given our assumptions concerning the cut-off mass scale
$M_{\rm cut}=M_{\rm fs}$ and also our adoption of the NFW profile for
these same haloes. In Appendix~\ref{app:mcut} and \ref{app:profiles},
we show that the predictions for the power spectrum are insensitive to
the precise value of $M_{\rm cut}$ (no changes are found over
two-orders of magnitude in mass around $M_{\rm fs}$), and the exact
shape of the profiles of these objects. More quantitatively, the
predictions are modified by $<1\%$ for scales $k<100\kMpc$ and for WDM
particle masses $m_{\rm WDM}>0.25~\keV$.


\begin{figure*}
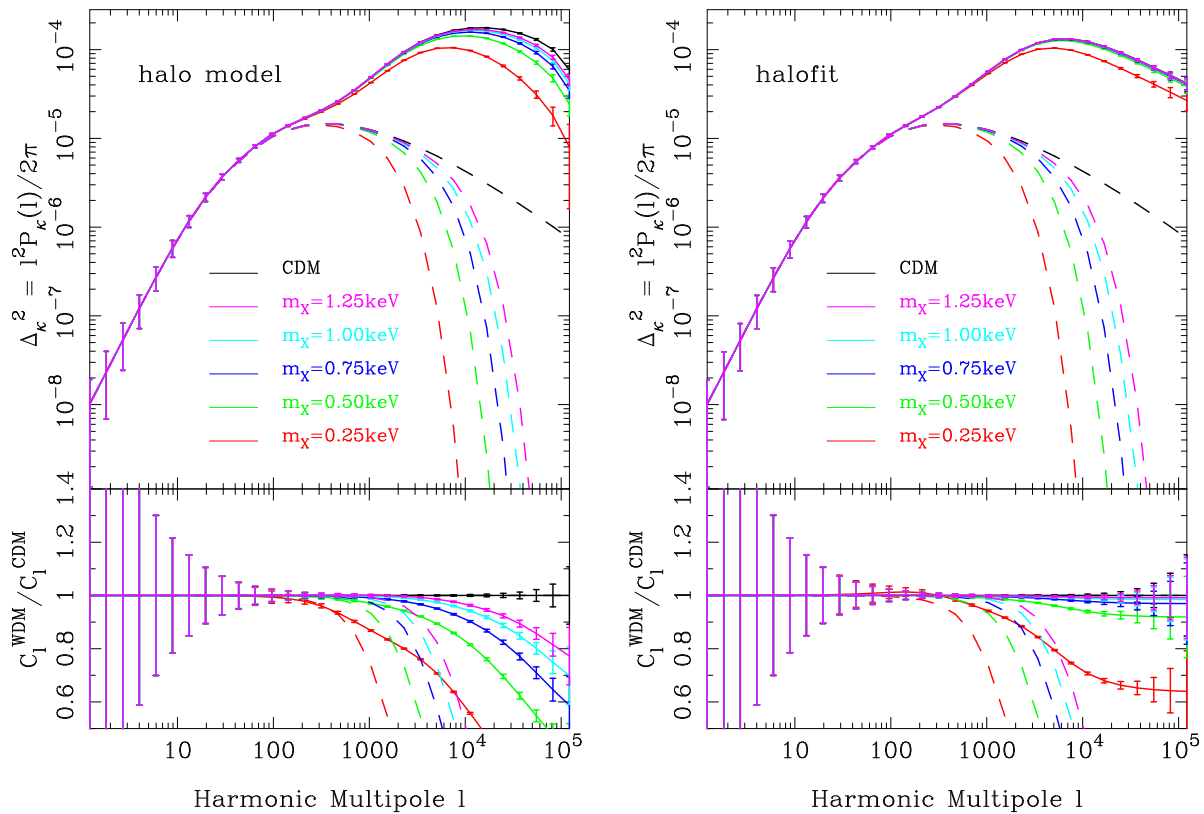

\centering{
  \includegraphics[width=7.6cm,clip=]{FIGS/Converg.HM.ps}\hspace{0.5cm}
  \includegraphics[width=7.6cm,clip=]{FIGS/Converg.halofit.ps}}
\caption{\small{Comparison of the weak lensing convergence power
    spectrum in the CDM and WDM models, as a function of angular
    multipole for source galaxies at $z_{\rm s}=1$. {\em Left and
    right panels}: predictions obtained using the halo model and {\tt
    halofit}, respectively. {\em Top panels:} Absolute power. Dashed
    and solid lines show the predictions from linear theory and from
    the nonlinear halo model. Black lines represent results for
    CDM. Colored lines with \{red, green, blue, cyan, magenta\}
    denote results for the WDM model with particle masses
    \mbox{$m_{\WDM}\equiv
    m_{\WDM}\in\{0.25,\,0.5,\,0.75,\,1.0,\,1.25\}\,{\rm keV}$}. {\em
    Bottom panel:} Ratio of the WDM model predictions to the CDM
    predictions. Lines styles and colors are as
    above.}\label{fig:converg}}
\end{figure*}


\subsection{Weak lensing convergence power spectra}

We now turn to the question: Can the modifications to the dark matter
model, of the kind considered in this paper, be detected with next
generation surveys? To answer this, we shall focus on weak
gravitational lensing by cosmic LSS as the most direct observable.

Weak lensing is the magnification and shearing of light rays from
distant source galaxies as they propagate through the inhomogeneous
matter in the Universe on their way towards the observer. Owing to our
ignorance of the intrinsic shapes of the unlensed galaxies, these
distortions can only be measured through the correlations in the
shapes of the source galaxies with each other. Following
\citep{BartelmannSchneider2001}, the weak lensing angular power
spectrum of convergence is given by
\be 
C_l = \frac{9H_0^4\Omega_{\rm m,0}^2}{4c^4}\int_{0}^{\chi_{\rm H}}
d\chi \frac{\overline{W}^2(\chi)}{a^2(\chi)}
P\left(k=\frac{l}{D_{A}(\chi)},\chi\right)\ ,
\ee
where $\chi$ is the comoving geodesic distance from the observer to
redshift $z$, and $D_{A}(\chi)$ is the comoving angular diameter
distance to redshift $z$. The survey weight function is defined:
\be 
\overline{W}(\chi)\equiv \int_{\chi}^{\chi_{\rm H}}d\chi' 
G(\chi') \frac{D_A(\chi'-\chi)}{D_{A}(\chi')}\label{eq:lensingweight} \ .
\ee
The function $G(\chi)$ is the normalized source distance distribution
function, which tells us the number of sources per unit distance
interval $d\chi$. For simplicity we shall assume the source
distribution is a Dirac delta function at a single source plane, hence
\mbox{$G(\chi)=\delta^{D}(\chi-\chi_{\rm s})$}. In this case the
survey weights simplify to \mbox{$\overline{W}(\chi)=D_A(\chi_{\rm
s}-\chi)/D_{A}(\chi_{\rm s})$} and we have,
\be 
C_l = \frac{9H_0^4\Omega_{\rm m,0}^2}{4c^4}\int_{0}^{\chi_{\rm s}} \frac{d\chi}{a^2(\chi)}
\left[\frac{D_A(\chi_{\rm s}-\chi)}{D_{A}(\chi_{\rm s})}\right]^2
P\left(\frac{l}{D_{A}(\chi)},\chi\right)\label{eq:convergOneSource} \ .
\ee
Direct numerical integration of the above expression can be very slow,
since it requires the evaluation of the halo model integrals for many
points in the $(k,a)$ plane, for a given $l$. The most efficient way
to overcome this problem, is to generate a bi-cubic spline
\citep{Pressetal1992} fit to $P(k,a)$ evaluated on a cubical grid of
$\log k$ and $\log a$. We found a mesh of size $100\times50$ gave good
results, with $k\in[0.005,100.0]$ and $a\in[0.5,1.0]$. Once the
bicubic spline is generated, the evaluation of the halo model $C_l$
takes of order the same time to compute as the linear $C_l$
calculation.

We now make a rough assessment of the expected errors on the
convergence power spectrum for a future weak lensing survey. To do
this, we assume that the spherical harmonic multipole coefficients are
independent Gaussianly distributed variables. In this case the
covariance matrix for a measurement of the power spectrum in band
powers $l$ and $l'$ can be written \citep{TakadaJain2004}:
\be {\rm Cov}\left[C_{l},C_{l'}\right]=\frac{1}{f_{\rm
    sky}}\frac{1}{N(l_m,l_n)}
\left[C_l+\frac{\sigma_{\gamma}^2}{\bar{n}}\right]^2\delta^{K}_{l,l'}\label{eq:ClErr}\ee
where $f_{\rm sky}$ is the fraction of sky covered, $\bar{n}$ is the
angular number density of galaxies on the sky, $\sigma_{\gamma}$ is
the shape noise error, which is the error on a given measurement of
the ellipticity of a galaxy. $N(l_m,l_n)$ is the number of independent
multipoles used to generate the estimate of $C_l$ in a given band
power. For a given band power $l\in[l_m,l_n]$, we find that 
\be
N(l_m,l_n) =  \sum_{l=l_{m}}^{l_n}\frac{2l+1}{2} 
=   \frac{\left(l_n+l_m+1\right)\left(l_n-l_m+1\right)}{2}\ .\ee
For our fiducial survey we assume future full-sky weak lensing
mission, like the proposed Euclid ~\citep{Euclid2010} or
LSST~\citep{LSST2009} surveys: $f_{\rm sky}=0.5$, $\bar{n}=35$
galaxies per square arc minute, $\sigma_{\gamma}=0.35$ and we take 50
bins in log-$l$ in the range $(1<l<10^5)$. 

The left panel of \Fig{fig:converg} shows the convergence matter power
spectrum for the linear and nonlinear halo model predictions for both
CDM and the WDM models. For large angular scales, $l<10^2$, the linear
and nonlinear predictions agree and there is no difference between CDM
and WDM. However, on smaller scales we see significant departures
between the predicted spectra. The importance of the nonlinear
corrections is very apparent: by $l\sim10^3$, there is an order of
magnitude difference between the linear and nonlinear predictions, and
by $l\sim10^4$, the difference is more than two orders of magnitude
for the WDM models. As expected, we also see that the measurement
errors are largest on very large and small scales, due to cosmic
variance and shape noise, respectively. 

The bottom panel of \Fig{fig:converg} shows the relative differences
between the WDM and CDM models. We see clearly that nonlinear
evolution reduces the differences between the WDM models and CDM.
However, the error bars show that the WDM models and the CDM should be
distinguishable at high significance.  

The right-hand panel of \Fig{fig:converg} shows the same but for the
case where the nonlinear power is computed using {\tt halofit}. The
main differences are: for $l\sim10^3$ the predictions from both
methods are roughly comparable, but for $l\gtrsim 10^{4}$ the halo
model has at least a factor 2 less signal than that obtained from {\tt
  halofit}.  The implications of this are that if the {\tt halofit} is
correct and the halo model is wrong, then this will diminish the
ability of future weak lensing surveys to constrain the mass of the
WDM particle. Nevertheless, as was shown recently in
\citep{Markovicetal2010}, reasonably good constraints may still be
achievable.  

Note that for these `signal-to-noise' predictions, we have assumed
that all of the source galaxies lie at $z_{\rm s}=1$. In the next
section we perform a more complete Fisher matrix analysis with a full
redshift distribution.


\begin{figure*}
\centering{
  \includegraphics[width=14cm,clip=]{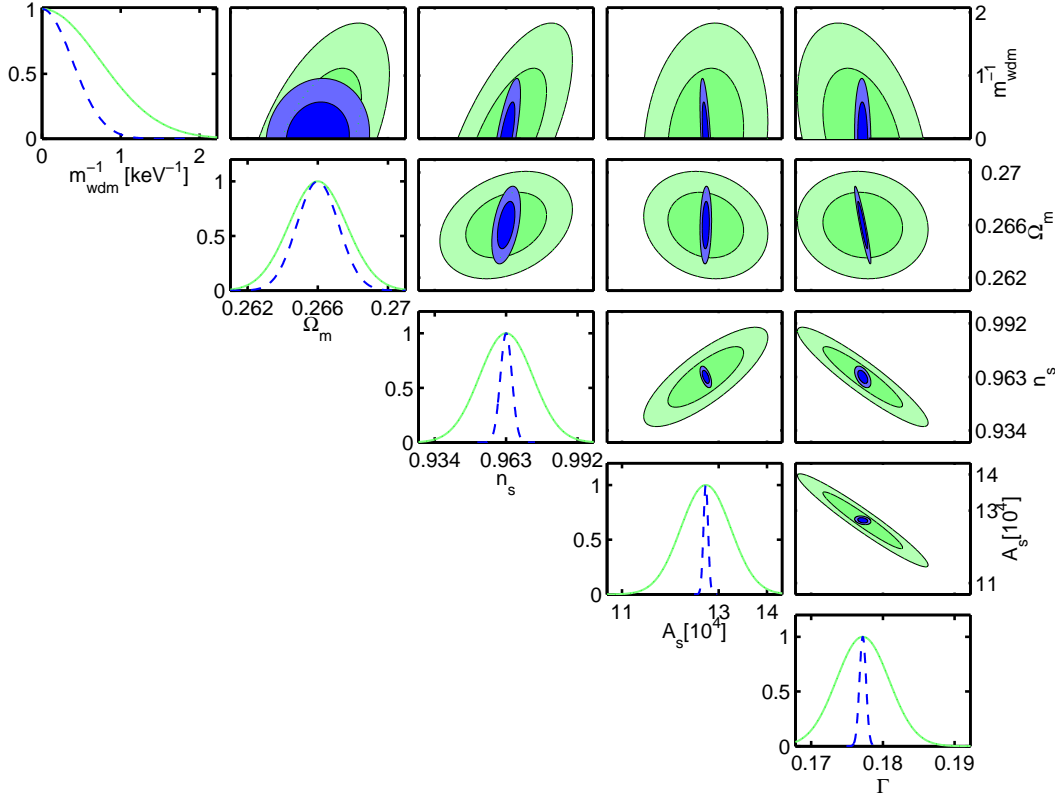}}
\caption{\small{We plot 65\% and 95\% likelihood contours,
    marginalised over all other parameters, for the fiducial
    (Euclid/LSST-like) survey (green) as well as the fiducial survey
    combined with Planck (blue). Our underlying fiducial model is the
    best fit WMAP7 $\Lambda$CDM model.}\label{fig:contours}}
\end{figure*}


\subsection{Fisher matrix forecast for $m_{\WDM}$}\label{sec:fisher}

Finally, we make a forecast for the lower limit on $m_{\WDM}$
obtainable with our fiducial (Euclid-like/LSST-like) survey. Owing to
these future missions including extensive photometric and
spectroscopic redshift survey components, we now take advantage of
this expected knowledge and assess the performance of a tomographic
lensing analysis \citep[see for example][]{Hu1999,TakadaJain2004}.

We adapt the analysis of the previous section as follows. We assume an
extended source redshift distribution for $G(z)$ in
\Eqn{eq:lensingweight}, and for this we take the same form as in
\citep{Markovicetal2010}.  We then divide the distribution into 10
separate tomographic redshift bins. The lensing convergence power
spectrum from the cross-correlation of sources in bins $i$ and $j$ is
then given by:
\be C_{(ij)}(l) = \int_{0}^{\chi_{\rm s}} 
\frac{d\chi}{a^2(\chi)} \widetilde{W}_i(\chi)\widetilde{W}_j(\chi) \chi^{-2}
P\left(\frac{l}{D_{A}(\chi)},\chi\right)\label{eq:converg} \ ,
\ee
where the weight function $\widetilde{W}_i(\chi)$ has the form:
\be \widetilde{W}_i(\chi) \equiv 
\left\{
\begin{array}{ll}
\frac{W_0}{\bar{n}a(\chi)}\chi \int_{\rm max(\chi,\chi_i)}^{\rm \chi_{i+1}} 
d\chi_s P_s(z)\frac{dz}{d\chi_s}\frac{\chi_s-\chi}{\chi_s} & (\chi\le\chi_{i+1})\\
0 & (\chi>\chi_{i+1})
\end{array}
\right. \ee
and where $W_0=3/2 \Omega_{m,0}H_0^2$ and $P_s(z)$ is the source
redshift distribution. Assuming that the underlying density field
obeys Gaussian statistics, then it can be shown that measurements of
the convergence power spectra for different tomographic bins are
correlated provided they share the same spherical harmonic multipole
$l$. In this case the covariance matrix can be written
\citep{TakadaJain2004}:
\ba
& & {\rm Cov}\left[C_{(ij)}^{\rm obs}(l),C_{(op)}^{\rm obs}(l')\right] =
\frac{1}{f_{\rm sky}}\frac{1}{N(l_m,l_n)}\delta^{K}_{l,l'}\nn \\
& & \hspace{1.5cm}\times \left[C_{(io)}^{\rm obs}(l)C_{(jp)}^{\rm obs}(l)+
C_{(ip)}^{\rm obs}(l)C_{(jo)}^{\rm obs}(l)\right]\label{eq:ClCov} \ , 
\label{eq:cov}\ea
where we defined $C_{(ij)}^{\rm obs}(l)\equiv
C_{(ij)}(l)+\delta_{(ij)}\sigma_{\gamma}^2/\bar{n}_{i}$. In the above
expression we have again ignored contributions to the covariance from
higher order spectra \citep{Smith2009}.

Assuming that the likelihood function for obtaining measurements of
the convergence power spectrum in several band powers and for several
tomographic bins is well described by a multivariate Gaussian
distribution with covariance matrix as given by \Eqn{eq:cov}, then 
the Fisher matrix can be written \citep{TakadaJain2004}:
\be {\mathcal F}_{\alpha\beta}=
\sum_l \sum_{X,Y}
\frac{\partial C_{(X)}(l)}{\partial\alpha} 
{\rm Cov}^{-1}\left[C_{(X)}^{\rm obs}(l),C_{(Y)}^{\rm obs}(l)\right]
\frac{C_{(X)}(l)}{\partial\beta}\ ,
\ee
where $\alpha$ and $\beta$ are elements of the vector of cosmological
model parameters upon which our measurements depend. $X$ and $Y$ label
the distinct pairs of tomographic bins, i.e.  \mbox{$X \in
  \{(1,1),\dots,(1,10),(2,2),\dots,(2,10),\dots\}$}.  The marginalized
variance and covariances of the cosmological parameters can be
obtained in the usual way \citep[for further details
  see][]{Heavens2009}: $\sigma^2_{\alpha\alpha}={\mathcal
  F}^{-1}_{\alpha\alpha}$ and $\sigma^2_{\alpha\beta}={\mathcal
  F}^{-1}_{\alpha\beta}$.

For our lensing analysis we choose to vary five parameters \mbox{${\bf
    p}\in\{m^{-1}_{\rm WDM},\Omega_{\rm m,0},n_{\rm s},A_{\rm
    s},\Gamma\}$}: the inverse WDM particle mass; the total matter
density today; the primordial perturbation spectral index; the
primordial spectral amplitude; and the matter power spectrum shape
parameter, $\Gamma\equiv e^{-2\Omega_{\rm b}h}\Omega_{\rm m}h$.  Note
that, since our fiducial model is $\Lambda$CDM with fiducial
parameters given by the WMAP7 data \citep{Komatsuetal2010}, we choose
to work with $m^{-1}_{\rm WDM}$ as a parameter in order to avoid
setting the fiducial model parameter to infinity. However, redefining
this variable does not avoid the strong variation of the error on
$m_{\rm WDM}$ with the chosen fiducial model. Furthermore, and most
problematically for such an analysis, the likelihood function is flat
with respect to the variation of this parameter around the maximum
likelihood point. For this reason we fit half a Gaussian curve on top
of this likelihood function and find single tailed errors as described
in \citet{Markovicetal2010}.

Figure~\ref{fig:contours} presents the 1- and 2-D likelihood contours
centered on our fiducial model. For the 2-D likelihood contours the 1-
and 2-$\sigma$ contours are denoted by the light and dark green shaded
regions, respectively. 

We then combine our lensing forecast with Planck
\citep{PlanckBlueBook} priors, i.e \mbox{${\mathcal
    F}_{\alpha\beta}^{\rm Total}={\mathcal F}_{\alpha\beta}^{\rm
    Lensing}+{\mathcal F}_{\alpha\beta}^{\rm Planck}$}. We use the
Planck Fisher matrix for the fiducial parameter set prescribed by
\citep{DETF2006}. We marginalize over the optical depth and the baryon
density and perform a parameter transformation as in
\citep{Eisensteinetal1999,Rassatetal2008}. As in
\citep{Markovicetal2010}, we assume that we will only have the 143 GHz
channel for CMB data. Furthermore, we assume that WDM with particle
masses of the order of a keV, does not leave an observable signature
in the CMB and hence set the corresponding entries in the Planck
Fisher matrix to zero.  The result of adding the Planck priors can be
seen in the \Fig{fig:contours} as the blue contours and lines.

Finally our forecast for the minimum limit on the WDM particle mass,
assuming that the CDM scenario is correct, from a Euclid/LSST-like
survey is $m_{\rm WDM}\ge 1.4 \keV$. On combining our fiducial lensing
survey and Planck data, the constraint tightens to $m_{\rm WDM} \ge
2.6 \keV$. We note that this limit is comparable to that found in
\citep{Markovicetal2010}. However, our analysis is more conservative
since we consider only multipoles $l\le 5000$, compared to $l\le
20,000$ used in the study by \citep{Markovicetal2010}. Moreover, at
the median redshift of the Euclid survey, $z_{\rm m}=0.9$, $l\le 5000$
corresponds to approximately $k_{\rm max}=1.6\kMpc$. The main reasons
for excluding small scales from the analysis are: one, to avoid the
range of scales where unknown baryonic physics effects inside clusters
may be a significant source of bias for interpreting the data
\citep{ZhanKnox2004}; two, the assumption of a decorrelated covariance
matrix for different $l$'s breaks down due to nonlinear mode coupling
in the density field, resulting in a loss of information
\cite{Scoccimarroetal2001,Satoetal2010}.


\section{Conclusions}\label{sec:conclusions}

In this paper we have explored a new approach to modelling the
nonlinear evolution of two-point mass clustering statistics in the WDM
scenario. The model is based on the phenomenological halo model
approach.

In \S\ref{sec:WDMtheory} we gave a brief overview of how linear
structure formation is changed in the WDM scenario. We defined a
free-streaming mass-scale $M_{\rm fs}$, below which halo formation is
likely to be strongly suppressed.
 
In \S\ref{sec:halomodel} we developed the halo model of large-scale
structure. We extended the model by including two components of
clustered mass in the Universe. That which is in collapsed dark matter
haloes and that which is in the form of smooth uncollapsed mass. This
extended model requires that we understand three correlation
functions: the halo-halo, smooth-halo and smooth-smooth correlation
functions. The halo-halo clustering can be calculated as in the
standard halo model. The clustering of the smooth matter is biased
with respect to the total mass. We showed that, provided one
understands the halo-halo correlation function, then the bias of the
smooth component can be derived from simple conservation arguments.

In \S\ref{sec:ingredients} we explored how the standard halo model
ingredients, i.e. halo mass function, halo bias and density profiles
are modified in the context of the WDM model. We showed that an
application of the standard mass function theory, using the WDM power
spectrum, leads to a 50\% suppression in halo abundance relative to
CDM for mass-scales $M\sim100M_{\rm fs}$. For our halo model
calculations we assumed that no haloes formed below $M_{\rm fs}$.

For the halo bias, we assumed that the standard peak-background split
calculations applied equally to the WDM model. Under this assumption,
we showed that the bias of haloes with $M<100M_{\rm fs}$ can be larger
by $\sim20\%$ for WDM models than CDM ones, for $m_{\WDM}=0.25\,
\keV$. Otherwise, halo bias is not affected. We also computed the bias
for the smooth component and found that it was anti-biased with
respect to the total mass, typically with a bias of order $b\sim0.7$,
for the range of WDM particle masses that we considered.

In modelling the impact of WDM on the density profiles, we took into
account two effects: firstly, the change in core density and
concentration due to the lower formation redshift in WDM models;
secondly, the suppression of density cusps due to relic thermal
velocities of the WDM particles. We found that the former effect is
the most important. Relic velocities only affect the density structure
on scales $r\lesssim 1\kpc$, these are irrelevant for weak lensing
analysis and so can be neglected.

In \S\ref{sec:results} we computed the nonlinear matter power spectrum
in the halo model and showed that, whilst the linear power is strongly
suppressed on small scales, nonlinear evolution generates significant
small-scale power. Relative to CDM the nonlinear WDM spectra are
suppressed by 10\% on scales $k\sim1\kMpc$ for a WDM particle with
mass $m_{\WDM}=0.25\,\keV$. For a heavier WDM particle the suppression
is much less, for $m_{\WDM}=1.0\,\keV$ the suppression is only
$\sim10\%$ on scales $k\sim20\kMpc$. We compared the predictions from
the halo model with those from the {\tt halofit} prescription
\citep{Smithetal2003}. We found that the latter showed more small
scale power, but that the differences between results for different
WDM particle masses was not as apparent as in the case for the Halo
Model. It will require high resolution $N$-body simulations to
validate which of the two models is more realistic.

We then computed the impact of a set of WDM particle masses on the
expected convergence power spectrum from a future weak lensing survey.
We also calculated the expected errors, assuming that each spherical
harmonic is an independent Gaussian variable. We found that, if our
halo model is correct, then it would be possible to differentiate
between different WDM candidates with a systematics free, full-sky,
weak lensing survey like Euclid \citep{Euclid2010} or LSST
\citep{LSST2009}.  

Finally, using the Fisher matrix approach we then forecasted the lower
limit on the WDM particle mass that would obtain from a Euclid like
weak-lensing mission. We made the conservative assumption that
multipoles $l<5000$ can only be safely interpreted, and we found the
lower limit $m_{\WDM}\ge 1.4\,\keV$ for Euclid only. On combining this
with a forecast of the Planck CMB data, the constraint tightened to
$m_{\WDM}\ge 2.6\,\keV$. Our findings are in agreement with the
earlier study of \cite{Markovicetal2010}, however these earlier
results were obtained assuming the less conservative multipole range
$l<20,000$.

CDM and WDM are equally plausible candidates for dark matter. We
should therefore invest resources in to future cosmological
experiments that allow us to place constraints on the mass of the dark
particle. We expect that a full-sky weak lensing survey combined with
the Planck data will greatly help achieve this objective.


\section*{Acknowledgements}
We thank Darren Reed for comments on an earlier draft, and an
anonymous referee for helpful suggestions. We thank Gianfranco
Bertone, Sarah Bridle, Alan Heavens, Martin Kilbinger, Ben Moore,
Aurel Schneider and Jochen Weller for useful discussions.  RES
acknowledges support from a Marie Curie Reintegration Grant, an award
for Experienced Researchers from the Alexander von Humboldt Foundation
and partial support from the Swiss National Foundation under contract
200021-116696/1.  KM acknowledges support from the International
Max-Planck Research School.



\input{HaloModel.WDM.4.bbl}

\appendix

\begin{figure}
\centering{
  \includegraphics[width=8.0cm,clip=]{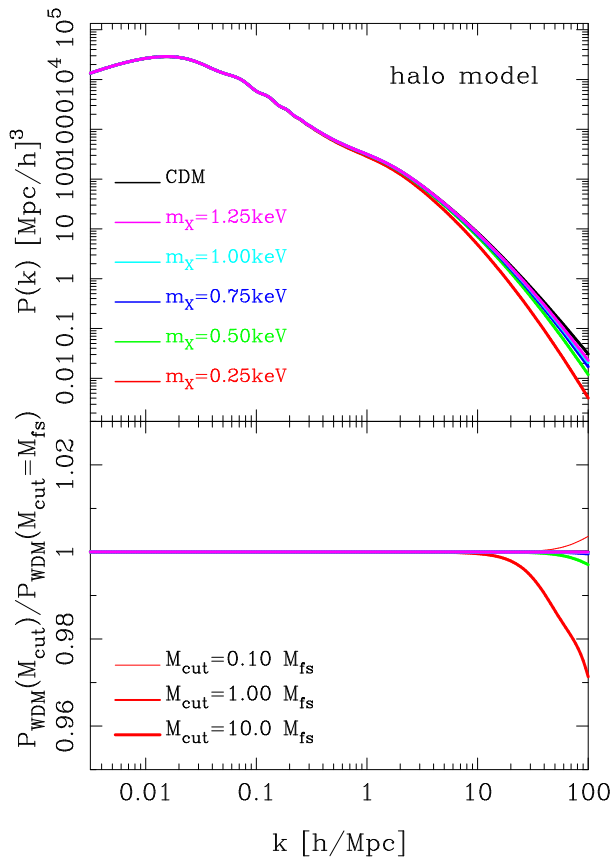}}
\caption{\small{Dependence of the halo model predictions for the WDM
    power spectra on the minimum mass scale $M_{\rm cut}$, as a
    function of wavenumber. {\em Top panel}: absolute power. Solid
    lines of increasing thickness denote the predictions with $M_{\rm
      cut}\in\{0.1,1.0,10.0\}M_{\rm fs}$, respectively.  Black lines
    denote CDM; and colors \{red, green, blue, cyan, magenta\} denote
    WDM particle masses \mbox{$m_{\WDM}\equiv
      m_{X}\in\{0.25,\,0.5,\,0.75,\,1.0,\,1.25\}\,{\rm keV}$}. {\em
      Bottom panels}: ratio of WDM power spectra with varying cut-off
    mass scales to the fiducial WDM predictions with $M_{\rm
      cut}=M_{\rm fs}$. 
  }\label{fig:PowMassCut}}
\end{figure}


\section{Testing the assumptions}

The halo model predictions hinge upon a number of assumptions.  We
shall now test some of these.

\subsection{Impact of Cut-off mass scale on predictions}
\label{app:mcut}

Perhaps, the most questionable assumption is that we assume only
haloes with $M>M_{\rm fs}$ exist and contribute to the clustering
strength on small scales. As was discussed earlier in
\S\ref{sec:WDMtheory} and \S\ref{sec:ingredientsI}, there are good
reasons for supposing that this is qualitatively correct. However, the
exact relation must be determined from numerical
simulations. Nevertheless, we may explore the importance of this
assumption for the predictions using our phenomenological model.

In the top panel of Figure~\ref{fig:PowMassCut} we show how the
nonlinear power spectra for our standard set of WDM models varies with
the cut-off mass scale. We consider three values for $M_{\rm
  cut}\in\{0.1,\,1.0,\,10.0\} M_{\rm fs}$, and in the figure these are
depicted as the continuous lines of the same colour but with
increasing thickness.  As can be seen, no differences are apparent for
the log-log plot.

The bottom panel of \Fig{fig:PowMassCut} shows the results in greater
detail, and here we present the fractional differences between the
power spectra with varying $M_{\rm cut}$ and our fiducial case where
$M_{\rm cut}=M_{\rm fs}$.  We find, for all of the scales and WDM
models considered, that the differences are $<1\%$. The one exception
is the $m_{\rm WDM}=0.25\,\keV$ model, for which we find that the
differences deviate by $\gtrsim1\%$ on scales $k\gtrsim30\kMpc$.

Owing to the fact that these tests have covered two orders of
magnitude in mass around the free-streaming mass scale, we are
therefore satisfied that the halo model predictions are robust to
changes in the fiducial value of $M_{\rm cut}$ adopted in this work.


\subsection{Importance of density profile shape for 
haloes at the cut-off mass scale}\label{app:profiles}

A further assumption that we have made is that all WDM haloes possess
NFW density profiles and that the characteristic density and hence
concentration parameters for these objects can be calculated as is
done for CDM. However, it might be argued that haloes with $M\gtrsim
M_{\rm cut}$, have profiles which are structurally substantially
different from the NFW model.  Of course, hi-resolution simulations
are needed to provide the definitive answer to this question. However,
we may use our phenomenological model to explore the importance of this
assumption on the predictions.

In fact, the analysis of the previous sub-section also helps clarify
matters here too. Let us suppose that the NFW model provides an
accurate description of haloes with $M\gg M_{\rm fs}$
\citep[][]{Avila-Reeseetal2001}, but for haloes with $M\gtrsim M_{\rm
  fs}$ that the profile is modified \citep{Colinetal2008}. Let us
consider the extreme case where the change to the halo profile in WDM
is so violent that the halo effectively has no profile. This might
simply be affected by removing the halo altogether from the
calculation. As we have shown in the previous sub-section, if we
add/remove haloes of mass one order of magnitude lower/higher than
$M_{\rm fs}$, the modifications to the predictions for the power
spectrum are $<1\%$ for $m_{\rm WDM}>0.25\,\keV$.

\vspace{0.2cm}

We thus conclude that the halo model predictions for the WDM power
spectrum, on scales $k<100\kMpc$, are insensitive ($<1\%$ changes) to
the exact value of the cut-off mass scale $M_{\rm cut}$ and the shapes
of the density profiles near to $M_{\rm fs}$, for $m_{\rm
  WDM}>0.25\,\keV$.


\end{document}

%% file: defs.tex
\newcommand\Eqn[1]     {Eq.\,(\ref{#1})}
\newcommand\Eqns[2]    {Eqs\,(\ref{#1}) and~(\ref{#2})}

\newcommand\Fig[1]     {Fig.\,{\ref{#1}}}

\newcommand\nn         {\nonumber}

\def\mnras{{Mon.~ Not.~ R.~ Astron.~ Soc.~}}

\def\prd{{Phys.~ Rev.~ D.~}}

\def\apj{{Astrophys.~ J.~}}
\def\apjs{{Astrophys.~ J.~ Suppl.~}}
\def\apjl{{Astrophys.~ J.~ Lett.~}}

\def\nat{{Nature (London)~}}

\def\mnras{{MNRAS}}

\def\prd{{PRD}}

\def\apj{{ApJ}}
\def\apjs{{ApJS}}
\def\apjl{{ApJL}}
\def\aap{{A\&A}}
\def\nat{{Nature}}
\def\aj{{Astronomical Journal}}

\def\physrep{{Phys.~ Rep.~}}
\def\jcap{Journal of Cosmology and Astro-Particle Physics}

\newcommand{\be}{\begin{equation}}
\newcommand{\ee}{\end{equation}}
\newcommand{\ba}{\begin{eqnarray}}
\newcommand{\ea}{\end{eqnarray}}

\def\NL{{\rm NL}}

\def\pp1{{\prime}}
\def\pp2{{\prime\prime}}

\def\2D{{\rm 2D}}

\def\h{{\rm h}}

\def\bx{{\bf x}}
\def\br{{\bf r}}

\def\bk{{\bf k}}

\def\by{{\bf y}}

\def\1Loop{{\rm 1Loop}}

\def\rhob{\bar{\rho}}

\def\Msol{h^{-1}M_{\odot}}
\def\kpc{\, h^{-1}{\rm kpc}}
\def\Mpc{\, h^{-1}{\rm Mpc}}

\def\kMpc{\, h \, {\rm Mpc}^{-1}}

\def\hh{{\rm hh}}

\def\dr{{\rm d}^3{\bf r}}
\def\dx{{\rm d}^3{\bf x}}
\def\dk{{\rm d}^3{\bf k}}

\font\BF=cmmib10

\def\by{{\hbox{\BF y}}}

\def\fun#1#2{\lower3.6pt\vbox{\baselineskip0pt\lineskip.9pt
        \ialign{$\mathsurround=0pt#1\hfill##\hfil$\crcr#2\crcr\sim\crcr}}}


%% file: HaloModel.WDM.4.bbl
\begin{thebibliography}{71}
\expandafter\ifx\csname natexlab\endcsname\relax\def\natexlab#1{#1}\fi
\expandafter\ifx\csname bibnamefont\endcsname\relax
  \def\bibnamefont#1{#1}\fi
\expandafter\ifx\csname bibfnamefont\endcsname\relax
  \def\bibfnamefont#1{#1}\fi
\expandafter\ifx\csname citenamefont\endcsname\relax
  \def\citenamefont#1{#1}\fi
\expandafter\ifx\csname url\endcsname\relax
  \def\url#1{\texttt{#1}}\fi
\expandafter\ifx\csname urlprefix\endcsname\relax\def\urlprefix{URL }\fi
\providecommand{\bibinfo}[2]{#2}
\providecommand{\eprint}[2][]{\url{#2}}

\bibitem[{\citenamefont{{Spergel} et~al.}(2007)\citenamefont{{Spergel}, {Bean},
  {Dor{\'e}}, {Nolta}, {Bennett}, {Dunkley}, {Hinshaw}, {Jarosik}, {Komatsu},
  {Page} et~al.}}]{Spergeletal2007}
\bibinfo{author}{\bibfnamefont{D.~N.} \bibnamefont{{Spergel}}},
  \bibinfo{author}{\bibfnamefont{R.}~\bibnamefont{{Bean}}},
  \bibinfo{author}{\bibfnamefont{O.}~\bibnamefont{{Dor{\'e}}}},
  \bibinfo{author}{\bibfnamefont{M.~R.} \bibnamefont{{Nolta}}},
  \bibinfo{author}{\bibfnamefont{C.~L.} \bibnamefont{{Bennett}}},
  \bibinfo{author}{\bibfnamefont{J.}~\bibnamefont{{Dunkley}}},
  \bibinfo{author}{\bibfnamefont{G.}~\bibnamefont{{Hinshaw}}},
  \bibinfo{author}{\bibfnamefont{N.}~\bibnamefont{{Jarosik}}},
  \bibinfo{author}{\bibfnamefont{E.}~\bibnamefont{{Komatsu}}},
  \bibinfo{author}{\bibfnamefont{L.}~\bibnamefont{{Page}}},
  \bibnamefont{et~al.}, \bibinfo{journal}{\apjs}
  \textbf{\bibinfo{volume}{170}}, \bibinfo{pages}{377} (\bibinfo{year}{2007}),
  \eprint{arXiv:astro-ph/0603449}.

\bibitem[{\citenamefont{{Bertone} et~al.}(2004)\citenamefont{{Bertone},
  {Hooper}, and {Silk}}}]{Bertoneetal2004}
\bibinfo{author}{\bibfnamefont{G.}~\bibnamefont{{Bertone}}},
  \bibinfo{author}{\bibfnamefont{D.}~\bibnamefont{{Hooper}}}, \bibnamefont{and}
  \bibinfo{author}{\bibfnamefont{J.}~\bibnamefont{{Silk}}},
  \bibinfo{journal}{\physrep} \textbf{\bibinfo{volume}{405}},
  \bibinfo{pages}{279} (\bibinfo{year}{2004}).

\bibitem[{\citenamefont{{Diemand} and {Moore}}(2009)}]{DiemandMoore2009}
\bibinfo{author}{\bibfnamefont{J.}~\bibnamefont{{Diemand}}} \bibnamefont{and}
  \bibinfo{author}{\bibfnamefont{B.}~\bibnamefont{{Moore}}},
  \bibinfo{journal}{ArXiv e-prints}  (\bibinfo{year}{2009}),
  \eprint{0906.4340}.

\bibitem[{\citenamefont{{Bertone}}(2010)}]{Bertone2010}
\bibinfo{editor}{\bibfnamefont{G.}~\bibnamefont{{Bertone}}}, ed.,
  \emph{\bibinfo{title}{{Particle Dark Matter: Observations, Models and
  Searches}}} (\bibinfo{publisher}{Cambridge University Press, Cambridge, UK},
  \bibinfo{year}{2010}).

\bibitem[{\citenamefont{{Moore}
  et~al.}(1999{\natexlab{a}})\citenamefont{{Moore}, {Ghigna}, {Governato},
  {Lake}, {Quinn}, {Stadel}, and {Tozzi}}}]{Mooreetal1999b}
\bibinfo{author}{\bibfnamefont{B.}~\bibnamefont{{Moore}}},
  \bibinfo{author}{\bibfnamefont{S.}~\bibnamefont{{Ghigna}}},
  \bibinfo{author}{\bibfnamefont{F.}~\bibnamefont{{Governato}}},
  \bibinfo{author}{\bibfnamefont{G.}~\bibnamefont{{Lake}}},
  \bibinfo{author}{\bibfnamefont{T.}~\bibnamefont{{Quinn}}},
  \bibinfo{author}{\bibfnamefont{J.}~\bibnamefont{{Stadel}}}, \bibnamefont{and}
  \bibinfo{author}{\bibfnamefont{P.}~\bibnamefont{{Tozzi}}},
  \bibinfo{journal}{\apjl} \textbf{\bibinfo{volume}{524}}, \bibinfo{pages}{L19}
  (\bibinfo{year}{1999}{\natexlab{a}}), \eprint{arXiv:astro-ph/9907411}.

\bibitem[{\citenamefont{{Klypin} et~al.}(1999)\citenamefont{{Klypin},
  {Kravtsov}, {Valenzuela}, and {Prada}}}]{Klypinetal1999}
\bibinfo{author}{\bibfnamefont{A.}~\bibnamefont{{Klypin}}},
  \bibinfo{author}{\bibfnamefont{A.~V.} \bibnamefont{{Kravtsov}}},
  \bibinfo{author}{\bibfnamefont{O.}~\bibnamefont{{Valenzuela}}},
  \bibnamefont{and} \bibinfo{author}{\bibfnamefont{F.}~\bibnamefont{{Prada}}},
  \bibinfo{journal}{\apj} \textbf{\bibinfo{volume}{522}}, \bibinfo{pages}{82}
  (\bibinfo{year}{1999}), \eprint{arXiv:astro-ph/9901240}.

\bibitem[{\citenamefont{{Peebles} and {Nusser}}(2010)}]{PeeblesNusser2010}
\bibinfo{author}{\bibfnamefont{P.~J.~E.} \bibnamefont{{Peebles}}}
  \bibnamefont{and} \bibinfo{author}{\bibfnamefont{A.}~\bibnamefont{{Nusser}}},
  \bibinfo{journal}{\nat} \textbf{\bibinfo{volume}{465}}, \bibinfo{pages}{565}
  (\bibinfo{year}{2010}), \eprint{1001.1484}.

\bibitem[{\citenamefont{{Diemand} et~al.}(2007)\citenamefont{{Diemand},
  {Kuhlen}, and {Madau}}}]{Diemandetal2007}
\bibinfo{author}{\bibfnamefont{J.}~\bibnamefont{{Diemand}}},
  \bibinfo{author}{\bibfnamefont{M.}~\bibnamefont{{Kuhlen}}}, \bibnamefont{and}
  \bibinfo{author}{\bibfnamefont{P.}~\bibnamefont{{Madau}}},
  \bibinfo{journal}{\apj} \textbf{\bibinfo{volume}{667}}, \bibinfo{pages}{859}
  (\bibinfo{year}{2007}), \eprint{arXiv:astro-ph/0703337}.

\bibitem[{\citenamefont{{Springel} et~al.}(2008)\citenamefont{{Springel},
  {Wang}, {Vogelsberger}, {Ludlow}, {Jenkins}, {Helmi}, {Navarro}, {Frenk}, and
  {White}}}]{Springeletal2008}
\bibinfo{author}{\bibfnamefont{V.}~\bibnamefont{{Springel}}},
  \bibinfo{author}{\bibfnamefont{J.}~\bibnamefont{{Wang}}},
  \bibinfo{author}{\bibfnamefont{M.}~\bibnamefont{{Vogelsberger}}},
  \bibinfo{author}{\bibfnamefont{A.}~\bibnamefont{{Ludlow}}},
  \bibinfo{author}{\bibfnamefont{A.}~\bibnamefont{{Jenkins}}},
  \bibinfo{author}{\bibfnamefont{A.}~\bibnamefont{{Helmi}}},
  \bibinfo{author}{\bibfnamefont{J.~F.} \bibnamefont{{Navarro}}},
  \bibinfo{author}{\bibfnamefont{C.~S.} \bibnamefont{{Frenk}}},
  \bibnamefont{and} \bibinfo{author}{\bibfnamefont{S.~D.~M.}
  \bibnamefont{{White}}}, \bibinfo{journal}{\mnras}
  \textbf{\bibinfo{volume}{391}}, \bibinfo{pages}{1685} (\bibinfo{year}{2008}),
  \eprint{0809.0898}.

\bibitem[{\citenamefont{{Zavala} et~al.}(2009)\citenamefont{{Zavala}, {Jing},
  {Faltenbacher}, {Yepes}, {Hoffman}, {Gottl{\"o}ber}, and
  {Catinella}}}]{Zavalaetal2009}
\bibinfo{author}{\bibfnamefont{J.}~\bibnamefont{{Zavala}}},
  \bibinfo{author}{\bibfnamefont{Y.~P.} \bibnamefont{{Jing}}},
  \bibinfo{author}{\bibfnamefont{A.}~\bibnamefont{{Faltenbacher}}},
  \bibinfo{author}{\bibfnamefont{G.}~\bibnamefont{{Yepes}}},
  \bibinfo{author}{\bibfnamefont{Y.}~\bibnamefont{{Hoffman}}},
  \bibinfo{author}{\bibfnamefont{S.}~\bibnamefont{{Gottl{\"o}ber}}},
  \bibnamefont{and}
  \bibinfo{author}{\bibfnamefont{B.}~\bibnamefont{{Catinella}}},
  \bibinfo{journal}{\apj} \textbf{\bibinfo{volume}{700}}, \bibinfo{pages}{1779}
  (\bibinfo{year}{2009}), \eprint{0906.0585}.

\bibitem[{\citenamefont{{Macci{\`o}} and
  {Fontanot}}(2010)}]{MaccioFontanot2010}
\bibinfo{author}{\bibfnamefont{A.~V.} \bibnamefont{{Macci{\`o}}}}
  \bibnamefont{and}
  \bibinfo{author}{\bibfnamefont{F.}~\bibnamefont{{Fontanot}}},
  \bibinfo{journal}{\mnras} \textbf{\bibinfo{volume}{404}},
  \bibinfo{pages}{L16} (\bibinfo{year}{2010}), \eprint{0910.2460}.

\bibitem[{\citenamefont{{Navarro} et~al.}(1997)\citenamefont{{Navarro},
  {Frenk}, and {White}}}]{Navarroetal1997}
\bibinfo{author}{\bibfnamefont{J.~F.} \bibnamefont{{Navarro}}},
  \bibinfo{author}{\bibfnamefont{C.~S.} \bibnamefont{{Frenk}}},
  \bibnamefont{and} \bibinfo{author}{\bibfnamefont{S.~D.~M.}
  \bibnamefont{{White}}}, \bibinfo{journal}{\apj}
  \textbf{\bibinfo{volume}{490}}, \bibinfo{pages}{493} (\bibinfo{year}{1997}),
  \eprint{arXiv:astro-ph/9611107}.

\bibitem[{\citenamefont{{Moore}
  et~al.}(1999{\natexlab{b}})\citenamefont{{Moore}, {Quinn}, {Governato},
  {Stadel}, and {Lake}}}]{Mooreetal1999c}
\bibinfo{author}{\bibfnamefont{B.}~\bibnamefont{{Moore}}},
  \bibinfo{author}{\bibfnamefont{T.}~\bibnamefont{{Quinn}}},
  \bibinfo{author}{\bibfnamefont{F.}~\bibnamefont{{Governato}}},
  \bibinfo{author}{\bibfnamefont{J.}~\bibnamefont{{Stadel}}}, \bibnamefont{and}
  \bibinfo{author}{\bibfnamefont{G.}~\bibnamefont{{Lake}}},
  \bibinfo{journal}{\mnras} \textbf{\bibinfo{volume}{310}},
  \bibinfo{pages}{1147} (\bibinfo{year}{1999}{\natexlab{b}}),
  \eprint{arXiv:astro-ph/9903164}.

\bibitem[{\citenamefont{{Ghigna} et~al.}(2000)\citenamefont{{Ghigna}, {Moore},
  {Governato}, {Lake}, {Quinn}, and {Stadel}}}]{Ghignaetal2000}
\bibinfo{author}{\bibfnamefont{S.}~\bibnamefont{{Ghigna}}},
  \bibinfo{author}{\bibfnamefont{B.}~\bibnamefont{{Moore}}},
  \bibinfo{author}{\bibfnamefont{F.}~\bibnamefont{{Governato}}},
  \bibinfo{author}{\bibfnamefont{G.}~\bibnamefont{{Lake}}},
  \bibinfo{author}{\bibfnamefont{T.}~\bibnamefont{{Quinn}}}, \bibnamefont{and}
  \bibinfo{author}{\bibfnamefont{J.}~\bibnamefont{{Stadel}}},
  \bibinfo{journal}{\apj} \textbf{\bibinfo{volume}{544}}, \bibinfo{pages}{616}
  (\bibinfo{year}{2000}), \eprint{arXiv:astro-ph/9910166}.

\bibitem[{\citenamefont{{Diemand} et~al.}(2004)\citenamefont{{Diemand},
  {Moore}, and {Stadel}}}]{Diemandetal2004}
\bibinfo{author}{\bibfnamefont{J.}~\bibnamefont{{Diemand}}},
  \bibinfo{author}{\bibfnamefont{B.}~\bibnamefont{{Moore}}}, \bibnamefont{and}
  \bibinfo{author}{\bibfnamefont{J.}~\bibnamefont{{Stadel}}},
  \bibinfo{journal}{\mnras} \textbf{\bibinfo{volume}{353}},
  \bibinfo{pages}{624} (\bibinfo{year}{2004}), \eprint{arXiv:astro-ph/0402267}.

\bibitem[{\citenamefont{{Stadel} et~al.}(2009)\citenamefont{{Stadel}, {Potter},
  {Moore}, {Diemand}, {Madau}, {Zemp}, {Kuhlen}, and
  {Quilis}}}]{Stadeletal2009}
\bibinfo{author}{\bibfnamefont{J.}~\bibnamefont{{Stadel}}},
  \bibinfo{author}{\bibfnamefont{D.}~\bibnamefont{{Potter}}},
  \bibinfo{author}{\bibfnamefont{B.}~\bibnamefont{{Moore}}},
  \bibinfo{author}{\bibfnamefont{J.}~\bibnamefont{{Diemand}}},
  \bibinfo{author}{\bibfnamefont{P.}~\bibnamefont{{Madau}}},
  \bibinfo{author}{\bibfnamefont{M.}~\bibnamefont{{Zemp}}},
  \bibinfo{author}{\bibfnamefont{M.}~\bibnamefont{{Kuhlen}}}, \bibnamefont{and}
  \bibinfo{author}{\bibfnamefont{V.}~\bibnamefont{{Quilis}}},
  \bibinfo{journal}{\mnras} \textbf{\bibinfo{volume}{398}},
  \bibinfo{pages}{L21} (\bibinfo{year}{2009}), \eprint{0808.2981}.

\bibitem[{\citenamefont{{Swaters} et~al.}(2003)\citenamefont{{Swaters},
  {Madore}, {van den Bosch}, and {Balcells}}}]{Swatersetal2003}
\bibinfo{author}{\bibfnamefont{R.~A.} \bibnamefont{{Swaters}}},
  \bibinfo{author}{\bibfnamefont{B.~F.} \bibnamefont{{Madore}}},
  \bibinfo{author}{\bibfnamefont{F.~C.} \bibnamefont{{van den Bosch}}},
  \bibnamefont{and}
  \bibinfo{author}{\bibfnamefont{M.}~\bibnamefont{{Balcells}}},
  \bibinfo{journal}{\apj} \textbf{\bibinfo{volume}{583}}, \bibinfo{pages}{732}
  (\bibinfo{year}{2003}), \eprint{arXiv:astro-ph/0210152}.

\bibitem[{\citenamefont{{Salucci} et~al.}(2007)\citenamefont{{Salucci}, {Lapi},
  {Tonini}, {Gentile}, {Yegorova}, and {Klein}}}]{Saluccietal2007}
\bibinfo{author}{\bibfnamefont{P.}~\bibnamefont{{Salucci}}},
  \bibinfo{author}{\bibfnamefont{A.}~\bibnamefont{{Lapi}}},
  \bibinfo{author}{\bibfnamefont{C.}~\bibnamefont{{Tonini}}},
  \bibinfo{author}{\bibfnamefont{G.}~\bibnamefont{{Gentile}}},
  \bibinfo{author}{\bibfnamefont{I.}~\bibnamefont{{Yegorova}}},
  \bibnamefont{and} \bibinfo{author}{\bibfnamefont{U.}~\bibnamefont{{Klein}}},
  \bibinfo{journal}{\mnras} \textbf{\bibinfo{volume}{378}}, \bibinfo{pages}{41}
  (\bibinfo{year}{2007}), \eprint{arXiv:astro-ph/0703115}.

\bibitem[{\citenamefont{{de Blok} et~al.}(2008)\citenamefont{{de Blok},
  {Walter}, {Brinks}, {Trachternach}, {Oh}, and {Kennicutt}}}]{deBloketal2008}
\bibinfo{author}{\bibfnamefont{W.~J.~G.} \bibnamefont{{de Blok}}},
  \bibinfo{author}{\bibfnamefont{F.}~\bibnamefont{{Walter}}},
  \bibinfo{author}{\bibfnamefont{E.}~\bibnamefont{{Brinks}}},
  \bibinfo{author}{\bibfnamefont{C.}~\bibnamefont{{Trachternach}}},
  \bibinfo{author}{\bibfnamefont{S.-H.} \bibnamefont{{Oh}}}, \bibnamefont{and}
  \bibinfo{author}{\bibfnamefont{R.~C.} \bibnamefont{{Kennicutt}},
  \bibfnamefont{Jr.}}, \bibinfo{journal}{\aj} \textbf{\bibinfo{volume}{136}},
  \bibinfo{pages}{2648} (\bibinfo{year}{2008}), \eprint{0810.2100}.

\bibitem[{\citenamefont{{Gentile} et~al.}(2009)\citenamefont{{Gentile},
  {Famaey}, {Zhao}, and {Salucci}}}]{Gentileetal2009}
\bibinfo{author}{\bibfnamefont{G.}~\bibnamefont{{Gentile}}},
  \bibinfo{author}{\bibfnamefont{B.}~\bibnamefont{{Famaey}}},
  \bibinfo{author}{\bibfnamefont{H.}~\bibnamefont{{Zhao}}}, \bibnamefont{and}
  \bibinfo{author}{\bibfnamefont{P.}~\bibnamefont{{Salucci}}},
  \bibinfo{journal}{\nat} \textbf{\bibinfo{volume}{461}}, \bibinfo{pages}{627}
  (\bibinfo{year}{2009}), \eprint{0909.5203}.

\bibitem[{\citenamefont{{Kroupa} et~al.}(2010)\citenamefont{{Kroupa}, {Famaey},
  {de Boer}, {Dabringhausen}, {Pawlowski}, {Boily}, {Jerjen}, {Forbes},
  {Hensler}, and {Metz}}}]{Kroupaetal2010}
\bibinfo{author}{\bibfnamefont{P.}~\bibnamefont{{Kroupa}}},
  \bibinfo{author}{\bibfnamefont{B.}~\bibnamefont{{Famaey}}},
  \bibinfo{author}{\bibfnamefont{K.~S.} \bibnamefont{{de Boer}}},
  \bibinfo{author}{\bibfnamefont{J.}~\bibnamefont{{Dabringhausen}}},
  \bibinfo{author}{\bibfnamefont{M.~S.} \bibnamefont{{Pawlowski}}},
  \bibinfo{author}{\bibfnamefont{C.~M.} \bibnamefont{{Boily}}},
  \bibinfo{author}{\bibfnamefont{H.}~\bibnamefont{{Jerjen}}},
  \bibinfo{author}{\bibfnamefont{D.}~\bibnamefont{{Forbes}}},
  \bibinfo{author}{\bibfnamefont{G.}~\bibnamefont{{Hensler}}},
  \bibnamefont{and} \bibinfo{author}{\bibfnamefont{M.}~\bibnamefont{{Metz}}},
  \bibinfo{journal}{\aap} \textbf{\bibinfo{volume}{523}}, \bibinfo{pages}{A32+}
  (\bibinfo{year}{2010}), \eprint{1006.1647}.

\bibitem[{\citenamefont{{Hogan} and {Dalcanton}}(2000)}]{HoganDalcanton2000}
\bibinfo{author}{\bibfnamefont{C.~J.} \bibnamefont{{Hogan}}} \bibnamefont{and}
  \bibinfo{author}{\bibfnamefont{J.~J.} \bibnamefont{{Dalcanton}}},
  \bibinfo{journal}{\prd} \textbf{\bibinfo{volume}{62}},
  \bibinfo{pages}{063511} (\bibinfo{year}{2000}),
  \eprint{arXiv:astro-ph/0002330}.

\bibitem[{\citenamefont{{Bode} et~al.}(2001)\citenamefont{{Bode}, {Ostriker},
  and {Turok}}}]{Bodeetal2001}
\bibinfo{author}{\bibfnamefont{P.}~\bibnamefont{{Bode}}},
  \bibinfo{author}{\bibfnamefont{J.~P.} \bibnamefont{{Ostriker}}},
  \bibnamefont{and} \bibinfo{author}{\bibfnamefont{N.}~\bibnamefont{{Turok}}},
  \bibinfo{journal}{\apj} \textbf{\bibinfo{volume}{556}}, \bibinfo{pages}{93}
  (\bibinfo{year}{2001}), \eprint{arXiv:astro-ph/0010389}.

\bibitem[{\citenamefont{{Boyanovsky}}(2010)}]{Boyanovsky2010}
\bibinfo{author}{\bibfnamefont{D.}~\bibnamefont{{Boyanovsky}}},
  \bibinfo{journal}{ArXiv e-prints}  (\bibinfo{year}{2010}),
  \eprint{1011.2217}.

\bibitem[{\citenamefont{{Ellis} et~al.}(1984)\citenamefont{{Ellis}, {Hagelin},
  {Nanopoulos}, {Olive}, and {Srednicki}}}]{Ellisetal1984}
\bibinfo{author}{\bibfnamefont{J.}~\bibnamefont{{Ellis}}},
  \bibinfo{author}{\bibfnamefont{J.~S.} \bibnamefont{{Hagelin}}},
  \bibinfo{author}{\bibfnamefont{D.~V.} \bibnamefont{{Nanopoulos}}},
  \bibinfo{author}{\bibfnamefont{K.}~\bibnamefont{{Olive}}}, \bibnamefont{and}
  \bibinfo{author}{\bibfnamefont{M.}~\bibnamefont{{Srednicki}}},
  \bibinfo{journal}{Nuclear Physics B} \textbf{\bibinfo{volume}{238}},
  \bibinfo{pages}{453} (\bibinfo{year}{1984}).

\bibitem[{\citenamefont{{Dodelson} and {Widrow}}(1994)}]{DodelsonWidrow1994}
\bibinfo{author}{\bibfnamefont{S.}~\bibnamefont{{Dodelson}}} \bibnamefont{and}
  \bibinfo{author}{\bibfnamefont{L.~M.} \bibnamefont{{Widrow}}},
  \bibinfo{journal}{Physical Review Letters} \textbf{\bibinfo{volume}{72}},
  \bibinfo{pages}{17} (\bibinfo{year}{1994}), \eprint{arXiv:hep-ph/9303287}.

\bibitem[{\citenamefont{{Boyarsky}
  et~al.}(2009{\natexlab{a}})\citenamefont{{Boyarsky}, {Lesgourgues},
  {Ruchayskiy}, and {Viel}}}]{Boyarskyetal2009a}
\bibinfo{author}{\bibfnamefont{A.}~\bibnamefont{{Boyarsky}}},
  \bibinfo{author}{\bibfnamefont{J.}~\bibnamefont{{Lesgourgues}}},
  \bibinfo{author}{\bibfnamefont{O.}~\bibnamefont{{Ruchayskiy}}},
  \bibnamefont{and} \bibinfo{author}{\bibfnamefont{M.}~\bibnamefont{{Viel}}},
  \bibinfo{journal}{\jcap} \textbf{\bibinfo{volume}{5}}, \bibinfo{pages}{12}
  (\bibinfo{year}{2009}{\natexlab{a}}), \eprint{0812.0010}.

\bibitem[{\citenamefont{{Col{\'{\i}}n}
  et~al.}(2000)\citenamefont{{Col{\'{\i}}n}, {Avila-Reese}, and
  {Valenzuela}}}]{Colinetal2000}
\bibinfo{author}{\bibfnamefont{P.}~\bibnamefont{{Col{\'{\i}}n}}},
  \bibinfo{author}{\bibfnamefont{V.}~\bibnamefont{{Avila-Reese}}},
  \bibnamefont{and}
  \bibinfo{author}{\bibfnamefont{O.}~\bibnamefont{{Valenzuela}}},
  \bibinfo{journal}{\apj} \textbf{\bibinfo{volume}{542}}, \bibinfo{pages}{622}
  (\bibinfo{year}{2000}), \eprint{arXiv:astro-ph/0004115}.

\bibitem[{\citenamefont{{White} and {Croft}}(2000)}]{WhiteCroft2000}
\bibinfo{author}{\bibfnamefont{M.}~\bibnamefont{{White}}} \bibnamefont{and}
  \bibinfo{author}{\bibfnamefont{R.~A.~C.} \bibnamefont{{Croft}}},
  \bibinfo{journal}{\apj} \textbf{\bibinfo{volume}{539}}, \bibinfo{pages}{497}
  (\bibinfo{year}{2000}), \eprint{arXiv:astro-ph/0001247}.

\bibitem[{\citenamefont{{Avila-Reese} et~al.}(2001)\citenamefont{{Avila-Reese},
  {Col{\'{\i}}n}, {Valenzuela}, {D'Onghia}, and
  {Firmani}}}]{Avila-Reeseetal2001}
\bibinfo{author}{\bibfnamefont{V.}~\bibnamefont{{Avila-Reese}}},
  \bibinfo{author}{\bibfnamefont{P.}~\bibnamefont{{Col{\'{\i}}n}}},
  \bibinfo{author}{\bibfnamefont{O.}~\bibnamefont{{Valenzuela}}},
  \bibinfo{author}{\bibfnamefont{E.}~\bibnamefont{{D'Onghia}}},
  \bibnamefont{and}
  \bibinfo{author}{\bibfnamefont{C.}~\bibnamefont{{Firmani}}},
  \bibinfo{journal}{\apj} \textbf{\bibinfo{volume}{559}}, \bibinfo{pages}{516}
  (\bibinfo{year}{2001}), \eprint{arXiv:astro-ph/0010525}.

\bibitem[{\citenamefont{{Bullock} et~al.}(2002)\citenamefont{{Bullock},
  {Kravtsov}, and {Col{\'{\i}}n}}}]{Bullocketal2002}
\bibinfo{author}{\bibfnamefont{J.~S.} \bibnamefont{{Bullock}}},
  \bibinfo{author}{\bibfnamefont{a.~A.~V.} \bibnamefont{{Kravtsov}}},
  \bibnamefont{and}
  \bibinfo{author}{\bibfnamefont{P.}~\bibnamefont{{Col{\'{\i}}n}}},
  \bibinfo{journal}{\apjl} \textbf{\bibinfo{volume}{564}}, \bibinfo{pages}{L1}
  (\bibinfo{year}{2002}), \eprint{arXiv:astro-ph/0109432}.

\bibitem[{\citenamefont{{Col{\'{\i}}n}
  et~al.}(2008)\citenamefont{{Col{\'{\i}}n}, {Valenzuela}, and
  {Avila-Reese}}}]{Colinetal2008}
\bibinfo{author}{\bibfnamefont{P.}~\bibnamefont{{Col{\'{\i}}n}}},
  \bibinfo{author}{\bibfnamefont{O.}~\bibnamefont{{Valenzuela}}},
  \bibnamefont{and}
  \bibinfo{author}{\bibfnamefont{V.}~\bibnamefont{{Avila-Reese}}},
  \bibinfo{journal}{\apj} \textbf{\bibinfo{volume}{673}}, \bibinfo{pages}{203}
  (\bibinfo{year}{2008}), \eprint{0709.4027}.

\bibitem[{\citenamefont{{Colombi} et~al.}(1996)\citenamefont{{Colombi},
  {Dodelson}, and {Widrow}}}]{Colombietal1996}
\bibinfo{author}{\bibfnamefont{S.}~\bibnamefont{{Colombi}}},
  \bibinfo{author}{\bibfnamefont{S.}~\bibnamefont{{Dodelson}}},
  \bibnamefont{and} \bibinfo{author}{\bibfnamefont{L.~M.}
  \bibnamefont{{Widrow}}}, \bibinfo{journal}{\apj}
  \textbf{\bibinfo{volume}{458}}, \bibinfo{pages}{1} (\bibinfo{year}{1996}),
  \eprint{arXiv:astro-ph/9505029}.

\bibitem[{\citenamefont{{Wang} and {White}}(2007)}]{WangWhite2007}
\bibinfo{author}{\bibfnamefont{J.}~\bibnamefont{{Wang}}} \bibnamefont{and}
  \bibinfo{author}{\bibfnamefont{S.~D.~M.} \bibnamefont{{White}}},
  \bibinfo{journal}{\mnras} \textbf{\bibinfo{volume}{380}}, \bibinfo{pages}{93}
  (\bibinfo{year}{2007}), \eprint{arXiv:astro-ph/0702575}.

\bibitem[{\citenamefont{{Polisensky} and
  {Ricotti}}(2010)}]{PolisenskyRicotti2010}
\bibinfo{author}{\bibfnamefont{E.}~\bibnamefont{{Polisensky}}}
  \bibnamefont{and}
  \bibinfo{author}{\bibfnamefont{M.}~\bibnamefont{{Ricotti}}},
  \bibinfo{journal}{ArXiv e-prints}  (\bibinfo{year}{2010}),
  \eprint{1004.1459}.

\bibitem[{\citenamefont{{Seljak} et~al.}(2006)\citenamefont{{Seljak},
  {Makarov}, {McDonald}, and {Trac}}}]{Seljaketal2006b}
\bibinfo{author}{\bibfnamefont{U.}~\bibnamefont{{Seljak}}},
  \bibinfo{author}{\bibfnamefont{A.}~\bibnamefont{{Makarov}}},
  \bibinfo{author}{\bibfnamefont{P.}~\bibnamefont{{McDonald}}},
  \bibnamefont{and} \bibinfo{author}{\bibfnamefont{H.}~\bibnamefont{{Trac}}},
  \bibinfo{journal}{Physical Review Letters} \textbf{\bibinfo{volume}{97}},
  \bibinfo{pages}{191303} (\bibinfo{year}{2006}),
  \eprint{arXiv:astro-ph/0602430}.

\bibitem[{\citenamefont{{Boyarsky} et~al.}(2008)\citenamefont{{Boyarsky},
  {Iakubovskyi}, {Ruchayskiy}, and {Savchenko}}}]{Boyarskyetal2008}
\bibinfo{author}{\bibfnamefont{A.}~\bibnamefont{{Boyarsky}}},
  \bibinfo{author}{\bibfnamefont{D.}~\bibnamefont{{Iakubovskyi}}},
  \bibinfo{author}{\bibfnamefont{O.}~\bibnamefont{{Ruchayskiy}}},
  \bibnamefont{and}
  \bibinfo{author}{\bibfnamefont{V.}~\bibnamefont{{Savchenko}}},
  \bibinfo{journal}{\mnras} \textbf{\bibinfo{volume}{387}},
  \bibinfo{pages}{1361} (\bibinfo{year}{2008}), \eprint{0709.2301}.

\bibitem[{\citenamefont{{Boyarsky}
  et~al.}(2009{\natexlab{b}})\citenamefont{{Boyarsky}, {Lesgourgues},
  {Ruchayskiy}, and {Viel}}}]{Boyarskyetal2009b}
\bibinfo{author}{\bibfnamefont{A.}~\bibnamefont{{Boyarsky}}},
  \bibinfo{author}{\bibfnamefont{J.}~\bibnamefont{{Lesgourgues}}},
  \bibinfo{author}{\bibfnamefont{O.}~\bibnamefont{{Ruchayskiy}}},
  \bibnamefont{and} \bibinfo{author}{\bibfnamefont{M.}~\bibnamefont{{Viel}}},
  \bibinfo{journal}{Physical Review Letters} \textbf{\bibinfo{volume}{102}},
  \bibinfo{pages}{201304} (\bibinfo{year}{2009}{\natexlab{b}}),
  \eprint{0812.3256}.

\bibitem[{\citenamefont{{Markovic} et~al.}(2010)\citenamefont{{Markovic},
  {Bridle}, {Slosar}, and {Weller}}}]{Markovicetal2010}
\bibinfo{author}{\bibfnamefont{K.}~\bibnamefont{{Markovic}}},
  \bibinfo{author}{\bibfnamefont{S.}~\bibnamefont{{Bridle}}},
  \bibinfo{author}{\bibfnamefont{A.}~\bibnamefont{{Slosar}}}, \bibnamefont{and}
  \bibinfo{author}{\bibfnamefont{J.}~\bibnamefont{{Weller}}},
  \bibinfo{journal}{ArXiv e-prints}  (\bibinfo{year}{2010}),
  \eprint{1009.0218}.

\bibitem[{\citenamefont{{Bond} et~al.}(1980)\citenamefont{{Bond}, {Efstathiou},
  and {Silk}}}]{Bondetal1980}
\bibinfo{author}{\bibfnamefont{J.~R.} \bibnamefont{{Bond}}},
  \bibinfo{author}{\bibfnamefont{G.}~\bibnamefont{{Efstathiou}}},
  \bibnamefont{and} \bibinfo{author}{\bibfnamefont{J.}~\bibnamefont{{Silk}}},
  \bibinfo{journal}{Physical Review Letters} \textbf{\bibinfo{volume}{45}},
  \bibinfo{pages}{1980} (\bibinfo{year}{1980}).

\bibitem[{\citenamefont{{Viel} et~al.}(2005)\citenamefont{{Viel},
  {Lesgourgues}, {Haehnelt}, {Matarrese}, and {Riotto}}}]{Vieletal2005}
\bibinfo{author}{\bibfnamefont{M.}~\bibnamefont{{Viel}}},
  \bibinfo{author}{\bibfnamefont{J.}~\bibnamefont{{Lesgourgues}}},
  \bibinfo{author}{\bibfnamefont{M.~G.} \bibnamefont{{Haehnelt}}},
  \bibinfo{author}{\bibfnamefont{S.}~\bibnamefont{{Matarrese}}},
  \bibnamefont{and} \bibinfo{author}{\bibfnamefont{A.}~\bibnamefont{{Riotto}}},
  \bibinfo{journal}{\prd} \textbf{\bibinfo{volume}{71}},
  \bibinfo{pages}{063534} (\bibinfo{year}{2005}),
  \eprint{arXiv:astro-ph/0501562}.

\bibitem[{\citenamefont{{Zentner} and {Bullock}}(2003)}]{ZentnerBullock2003}
\bibinfo{author}{\bibfnamefont{A.~R.} \bibnamefont{{Zentner}}}
  \bibnamefont{and} \bibinfo{author}{\bibfnamefont{J.~S.}
  \bibnamefont{{Bullock}}}, \bibinfo{journal}{\apj}
  \textbf{\bibinfo{volume}{598}}, \bibinfo{pages}{49} (\bibinfo{year}{2003}),
  \eprint{arXiv:astro-ph/0304292}.

\bibitem[{\citenamefont{{Cooray} and {Sheth}}(2002)}]{CooraySheth2002}
\bibinfo{author}{\bibfnamefont{A.}~\bibnamefont{{Cooray}}} \bibnamefont{and}
  \bibinfo{author}{\bibfnamefont{R.}~\bibnamefont{{Sheth}}},
  \bibinfo{journal}{\physrep} \textbf{\bibinfo{volume}{372}},
  \bibinfo{pages}{1} (\bibinfo{year}{2002}), \eprint{arXiv:astro-ph/0206508}.

\bibitem[{\citenamefont{{Smith} et~al.}(2007)\citenamefont{{Smith},
  {Scoccimarro}, and {Sheth}}}]{Smithetal2007}
\bibinfo{author}{\bibfnamefont{R.~E.} \bibnamefont{{Smith}}},
  \bibinfo{author}{\bibfnamefont{R.}~\bibnamefont{{Scoccimarro}}},
  \bibnamefont{and} \bibinfo{author}{\bibfnamefont{R.~K.}
  \bibnamefont{{Sheth}}}, \bibinfo{journal}{\prd}
  \textbf{\bibinfo{volume}{75}}, \bibinfo{pages}{063512}
  (\bibinfo{year}{2007}), \eprint{arXiv:astro-ph/0609547}.

\bibitem[{\citenamefont{{Fry} and {Gaztanaga}}(1993)}]{FryGaztanaga1993}
\bibinfo{author}{\bibfnamefont{J.~N.} \bibnamefont{{Fry}}} \bibnamefont{and}
  \bibinfo{author}{\bibfnamefont{E.}~\bibnamefont{{Gaztanaga}}},
  \bibinfo{journal}{\apj} \textbf{\bibinfo{volume}{413}}, \bibinfo{pages}{447}
  (\bibinfo{year}{1993}), \eprint{arXiv:astro-ph/9302009}.

\bibitem[{\citenamefont{{Mo} and {White}}(1996)}]{MoWhite1996}
\bibinfo{author}{\bibfnamefont{H.~J.} \bibnamefont{{Mo}}} \bibnamefont{and}
  \bibinfo{author}{\bibfnamefont{S.~D.~M.} \bibnamefont{{White}}},
  \bibinfo{journal}{\mnras} \textbf{\bibinfo{volume}{282}},
  \bibinfo{pages}{347} (\bibinfo{year}{1996}), \eprint{arXiv:astro-ph/9512127}.

\bibitem[{\citenamefont{{Takada} and {Jain}}(2004)}]{TakadaJain2004}
\bibinfo{author}{\bibfnamefont{M.}~\bibnamefont{{Takada}}} \bibnamefont{and}
  \bibinfo{author}{\bibfnamefont{B.}~\bibnamefont{{Jain}}},
  \bibinfo{journal}{\mnras} \textbf{\bibinfo{volume}{348}},
  \bibinfo{pages}{897} (\bibinfo{year}{2004}), \eprint{arXiv:astro-ph/0310125}.

\bibitem[{\citenamefont{{Smith} et~al.}(2010)\citenamefont{{Smith},
  {Desjacques}, and {Marian}}}]{Smithetal2010}
\bibinfo{author}{\bibfnamefont{R.~E.} \bibnamefont{{Smith}}},
  \bibinfo{author}{\bibfnamefont{V.}~\bibnamefont{{Desjacques}}},
  \bibnamefont{and} \bibinfo{author}{\bibfnamefont{L.}~\bibnamefont{{Marian}}},
  \bibinfo{journal}{ArXiv e-prints}  (\bibinfo{year}{2010}),
  \eprint{1009.5085}.

\bibitem[{\citenamefont{{Peebles}}(1980)}]{Peebles1980}
\bibinfo{author}{\bibfnamefont{P.~J.~E.} \bibnamefont{{Peebles}}},
  \emph{\bibinfo{title}{{The large-scale structure of the universe}}}
  (\bibinfo{publisher}{Research supported by the National Science
  Foundation.~Princeton, N.J., Princeton University Press, 1980.~435 p.},
  \bibinfo{year}{1980}).

\bibitem[{\citenamefont{{Peacock}}(1999)}]{Peacock1999}
\bibinfo{author}{\bibfnamefont{J.~A.} \bibnamefont{{Peacock}}},
  \emph{\bibinfo{title}{{Cosmological Physics}}}
  (\bibinfo{publisher}{Cambridge, UK: Cambridge University Press, 1999.},
  \bibinfo{year}{1999}).

\bibitem[{\citenamefont{{Press} and {Schechter}}(1974)}]{PressSchechter1974}
\bibinfo{author}{\bibfnamefont{W.~H.} \bibnamefont{{Press}}} \bibnamefont{and}
  \bibinfo{author}{\bibfnamefont{P.}~\bibnamefont{{Schechter}}},
  \bibinfo{journal}{\apj} \textbf{\bibinfo{volume}{187}}, \bibinfo{pages}{425}
  (\bibinfo{year}{1974}).

\bibitem[{\citenamefont{{Sheth} and {Tormen}}(1999)}]{ShethTormen1999}
\bibinfo{author}{\bibfnamefont{R.~K.} \bibnamefont{{Sheth}}} \bibnamefont{and}
  \bibinfo{author}{\bibfnamefont{G.}~\bibnamefont{{Tormen}}},
  \bibinfo{journal}{\mnras} \textbf{\bibinfo{volume}{308}},
  \bibinfo{pages}{119} (\bibinfo{year}{1999}), \eprint{arXiv:astro-ph/9901122}.

\bibitem[{\citenamefont{{Bullock} et~al.}(2001)\citenamefont{{Bullock},
  {Kolatt}, {Sigad}, {Somerville}, {Kravtsov}, {Klypin}, {Primack}, and
  {Dekel}}}]{Bullocketal2001}
\bibinfo{author}{\bibfnamefont{J.~S.} \bibnamefont{{Bullock}}},
  \bibinfo{author}{\bibfnamefont{T.~S.} \bibnamefont{{Kolatt}}},
  \bibinfo{author}{\bibfnamefont{Y.}~\bibnamefont{{Sigad}}},
  \bibinfo{author}{\bibfnamefont{R.~S.} \bibnamefont{{Somerville}}},
  \bibinfo{author}{\bibfnamefont{A.~V.} \bibnamefont{{Kravtsov}}},
  \bibinfo{author}{\bibfnamefont{A.~A.} \bibnamefont{{Klypin}}},
  \bibinfo{author}{\bibfnamefont{J.~R.} \bibnamefont{{Primack}}},
  \bibnamefont{and} \bibinfo{author}{\bibfnamefont{A.}~\bibnamefont{{Dekel}}},
  \bibinfo{journal}{\mnras} \textbf{\bibinfo{volume}{321}},
  \bibinfo{pages}{559} (\bibinfo{year}{2001}), \eprint{arXiv:astro-ph/9908159}.

\bibitem[{\citenamefont{{Smith} and {Watts}}(2005)}]{SmithWatts2005}
\bibinfo{author}{\bibfnamefont{R.~E.} \bibnamefont{{Smith}}} \bibnamefont{and}
  \bibinfo{author}{\bibfnamefont{P.~I.~R.} \bibnamefont{{Watts}}},
  \bibinfo{journal}{\mnras} \textbf{\bibinfo{volume}{360}},
  \bibinfo{pages}{203} (\bibinfo{year}{2005}), \eprint{arXiv:astro-ph/0412441}.

\bibitem[{\citenamefont{{Tremaine} and {Gunn}}(1979)}]{TremaineGunn1979}
\bibinfo{author}{\bibfnamefont{S.}~\bibnamefont{{Tremaine}}} \bibnamefont{and}
  \bibinfo{author}{\bibfnamefont{J.~E.} \bibnamefont{{Gunn}}},
  \bibinfo{journal}{Physical Review Letters} \textbf{\bibinfo{volume}{42}},
  \bibinfo{pages}{407} (\bibinfo{year}{1979}).

\bibitem[{\citenamefont{{Smith} et~al.}(2003)\citenamefont{{Smith}, {Peacock},
  {Jenkins}, {White}, {Frenk}, {Pearce}, {Thomas}, {Efstathiou}, and
  {Couchman}}}]{Smithetal2003}
\bibinfo{author}{\bibfnamefont{R.~E.} \bibnamefont{{Smith}}},
  \bibinfo{author}{\bibfnamefont{J.~A.} \bibnamefont{{Peacock}}},
  \bibinfo{author}{\bibfnamefont{A.}~\bibnamefont{{Jenkins}}},
  \bibinfo{author}{\bibfnamefont{S.~D.~M.} \bibnamefont{{White}}},
  \bibinfo{author}{\bibfnamefont{C.~S.} \bibnamefont{{Frenk}}},
  \bibinfo{author}{\bibfnamefont{F.~R.} \bibnamefont{{Pearce}}},
  \bibinfo{author}{\bibfnamefont{P.~A.} \bibnamefont{{Thomas}}},
  \bibinfo{author}{\bibfnamefont{G.}~\bibnamefont{{Efstathiou}}},
  \bibnamefont{and} \bibinfo{author}{\bibfnamefont{H.~M.~P.}
  \bibnamefont{{Couchman}}}, \bibinfo{journal}{\mnras}
  \textbf{\bibinfo{volume}{341}}, \bibinfo{pages}{1311} (\bibinfo{year}{2003}),
  \eprint{arXiv:astro-ph/0207664}.

\bibitem[{\citenamefont{{Bartelmann} and
  {Schneider}}(2001)}]{BartelmannSchneider2001}
\bibinfo{author}{\bibfnamefont{M.}~\bibnamefont{{Bartelmann}}}
  \bibnamefont{and}
  \bibinfo{author}{\bibfnamefont{P.}~\bibnamefont{{Schneider}}},
  \bibinfo{journal}{\physrep} \textbf{\bibinfo{volume}{340}},
  \bibinfo{pages}{291} (\bibinfo{year}{2001}), \eprint{arXiv:astro-ph/9912508}.

\bibitem[{\citenamefont{{Press} et~al.}(1992)\citenamefont{{Press},
  {Teukolsky}, {Vetterling}, and {Flannery}}}]{Pressetal1992}
\bibinfo{author}{\bibfnamefont{W.~H.} \bibnamefont{{Press}}},
  \bibinfo{author}{\bibfnamefont{S.~A.} \bibnamefont{{Teukolsky}}},
  \bibinfo{author}{\bibfnamefont{W.~T.} \bibnamefont{{Vetterling}}},
  \bibnamefont{and} \bibinfo{author}{\bibfnamefont{B.~P.}
  \bibnamefont{{Flannery}}}, \emph{\bibinfo{title}{{Numerical recipes in
  FORTRAN. The art of scientific computing}}} (\bibinfo{publisher}{Cambridge:
  University Press, |c1992, 2nd ed.}, \bibinfo{year}{1992}).

\bibitem[{\citenamefont{{Refregier} et~al.}(2010)\citenamefont{{Refregier},
  {Amara}, {Kitching}, {Rassat}, {Scaramella}, {Weller}, and {Euclid Imaging
  Consortium}}}]{Euclid2010}
\bibinfo{author}{\bibfnamefont{A.}~\bibnamefont{{Refregier}}},
  \bibinfo{author}{\bibfnamefont{A.}~\bibnamefont{{Amara}}},
  \bibinfo{author}{\bibfnamefont{T.~D.} \bibnamefont{{Kitching}}},
  \bibinfo{author}{\bibfnamefont{A.}~\bibnamefont{{Rassat}}},
  \bibinfo{author}{\bibfnamefont{R.}~\bibnamefont{{Scaramella}}},
  \bibinfo{author}{\bibfnamefont{J.}~\bibnamefont{{Weller}}}, \bibnamefont{and}
  \bibinfo{author}{\bibfnamefont{f.~t.} \bibnamefont{{Euclid Imaging
  Consortium}}}, \bibinfo{journal}{ArXiv e-prints}  (\bibinfo{year}{2010}),
  \eprint{1001.0061}.

\bibitem[{\citenamefont{{LSST}}(2009)}]{LSST2009}
\bibinfo{author}{\bibnamefont{{LSST}}}, \bibinfo{journal}{ArXiv e-prints}
  (\bibinfo{year}{2009}), \eprint{0912.0201}.

\bibitem[{\citenamefont{{Hu}}(1999)}]{Hu1999}
\bibinfo{author}{\bibfnamefont{W.}~\bibnamefont{{Hu}}},
  \bibinfo{journal}{\apjl} \textbf{\bibinfo{volume}{522}}, \bibinfo{pages}{L21}
  (\bibinfo{year}{1999}), \eprint{arXiv:astro-ph/9904153}.

\bibitem[{\citenamefont{{Smith}}(2009)}]{Smith2009}
\bibinfo{author}{\bibfnamefont{R.~E.} \bibnamefont{{Smith}}},
  \bibinfo{journal}{\mnras} pp. \bibinfo{pages}{1337--+}
  (\bibinfo{year}{2009}), \eprint{0810.1960}.

\bibitem[{\citenamefont{{Heavens}}(2009)}]{Heavens2009}
\bibinfo{author}{\bibfnamefont{A.}~\bibnamefont{{Heavens}}},
  \bibinfo{journal}{ArXiv e-prints}  (\bibinfo{year}{2009}),
  \eprint{0906.0664}.

\bibitem[{\citenamefont{{Komatsu} et~al.}(2010)\citenamefont{{Komatsu},
  {Smith}, {Dunkley}, {Bennett}, {Gold}, {Hinshaw}, {Jarosik}, {Larson},
  {Nolta}, {Page} et~al.}}]{Komatsuetal2010}
\bibinfo{author}{\bibfnamefont{E.}~\bibnamefont{{Komatsu}}},
  \bibinfo{author}{\bibfnamefont{K.~M.} \bibnamefont{{Smith}}},
  \bibinfo{author}{\bibfnamefont{J.}~\bibnamefont{{Dunkley}}},
  \bibinfo{author}{\bibfnamefont{C.~L.} \bibnamefont{{Bennett}}},
  \bibinfo{author}{\bibfnamefont{B.}~\bibnamefont{{Gold}}},
  \bibinfo{author}{\bibfnamefont{G.}~\bibnamefont{{Hinshaw}}},
  \bibinfo{author}{\bibfnamefont{N.}~\bibnamefont{{Jarosik}}},
  \bibinfo{author}{\bibfnamefont{D.}~\bibnamefont{{Larson}}},
  \bibinfo{author}{\bibfnamefont{M.~R.} \bibnamefont{{Nolta}}},
  \bibinfo{author}{\bibfnamefont{L.}~\bibnamefont{{Page}}},
  \bibnamefont{et~al.}, \bibinfo{journal}{ArXiv e-prints}
  (\bibinfo{year}{2010}), \eprint{1001.4538}.

\bibitem[{\citenamefont{{The Planck Collaboration}}(2006)}]{PlanckBlueBook}
\bibinfo{author}{\bibnamefont{{The Planck Collaboration}}},
  \bibinfo{journal}{ArXiv Astrophysics e-prints}  (\bibinfo{year}{2006}),
  \eprint{arXiv:astro-ph/0604069}.

\bibitem[{\citenamefont{{Albrecht} et~al.}(2006)\citenamefont{{Albrecht},
  {Bernstein}, {Cahn}, {Freedman}, {Hewitt}, {Hu}, {Huth}, {Kamionkowski},
  {Kolb}, {Knox} et~al.}}]{DETF2006}
\bibinfo{author}{\bibfnamefont{A.}~\bibnamefont{{Albrecht}}},
  \bibinfo{author}{\bibfnamefont{G.}~\bibnamefont{{Bernstein}}},
  \bibinfo{author}{\bibfnamefont{R.}~\bibnamefont{{Cahn}}},
  \bibinfo{author}{\bibfnamefont{W.~L.} \bibnamefont{{Freedman}}},
  \bibinfo{author}{\bibfnamefont{J.}~\bibnamefont{{Hewitt}}},
  \bibinfo{author}{\bibfnamefont{W.}~\bibnamefont{{Hu}}},
  \bibinfo{author}{\bibfnamefont{J.}~\bibnamefont{{Huth}}},
  \bibinfo{author}{\bibfnamefont{M.}~\bibnamefont{{Kamionkowski}}},
  \bibinfo{author}{\bibfnamefont{E.~W.} \bibnamefont{{Kolb}}},
  \bibinfo{author}{\bibfnamefont{L.}~\bibnamefont{{Knox}}},
  \bibnamefont{et~al.}, \bibinfo{journal}{ArXiv Astrophysics e-prints}
  (\bibinfo{year}{2006}), \eprint{arXiv:astro-ph/0609591}.

\bibitem[{\citenamefont{{Eisenstein} et~al.}(1999)\citenamefont{{Eisenstein},
  {Hu}, and {Tegmark}}}]{Eisensteinetal1999}
\bibinfo{author}{\bibfnamefont{D.~J.} \bibnamefont{{Eisenstein}}},
  \bibinfo{author}{\bibfnamefont{W.}~\bibnamefont{{Hu}}}, \bibnamefont{and}
  \bibinfo{author}{\bibfnamefont{M.}~\bibnamefont{{Tegmark}}},
  \bibinfo{journal}{\apj} \textbf{\bibinfo{volume}{518}}, \bibinfo{pages}{2}
  (\bibinfo{year}{1999}), \eprint{arXiv:astro-ph/9807130}.

\bibitem[{\citenamefont{{Rassat} et~al.}(2008)\citenamefont{{Rassat}, {Amara},
  {Amendola}, {Castander}, {Kitching}, {Kunz}, {Refregier}, {Wang}, and
  {Weller}}}]{Rassatetal2008}
\bibinfo{author}{\bibfnamefont{A.}~\bibnamefont{{Rassat}}},
  \bibinfo{author}{\bibfnamefont{A.}~\bibnamefont{{Amara}}},
  \bibinfo{author}{\bibfnamefont{L.}~\bibnamefont{{Amendola}}},
  \bibinfo{author}{\bibfnamefont{F.~J.} \bibnamefont{{Castander}}},
  \bibinfo{author}{\bibfnamefont{T.}~\bibnamefont{{Kitching}}},
  \bibinfo{author}{\bibfnamefont{M.}~\bibnamefont{{Kunz}}},
  \bibinfo{author}{\bibfnamefont{A.}~\bibnamefont{{Refregier}}},
  \bibinfo{author}{\bibfnamefont{Y.}~\bibnamefont{{Wang}}}, \bibnamefont{and}
  \bibinfo{author}{\bibfnamefont{J.}~\bibnamefont{{Weller}}},
  \bibinfo{journal}{ArXiv e-prints}  (\bibinfo{year}{2008}),
  \eprint{0810.0003}.

\bibitem[{\citenamefont{{Zhan} and {Knox}}(2004)}]{ZhanKnox2004}
\bibinfo{author}{\bibfnamefont{H.}~\bibnamefont{{Zhan}}} \bibnamefont{and}
  \bibinfo{author}{\bibfnamefont{L.}~\bibnamefont{{Knox}}},
  \bibinfo{journal}{\apjl} \textbf{\bibinfo{volume}{616}}, \bibinfo{pages}{L75}
  (\bibinfo{year}{2004}), \eprint{arXiv:astro-ph/0409198}.

\bibitem[{\citenamefont{{Scoccimarro} et~al.}(2001)\citenamefont{{Scoccimarro},
  {Sheth}, {Hui}, and {Jain}}}]{Scoccimarroetal2001}
\bibinfo{author}{\bibfnamefont{R.}~\bibnamefont{{Scoccimarro}}},
  \bibinfo{author}{\bibfnamefont{R.~K.} \bibnamefont{{Sheth}}},
  \bibinfo{author}{\bibfnamefont{L.}~\bibnamefont{{Hui}}}, \bibnamefont{and}
  \bibinfo{author}{\bibfnamefont{B.}~\bibnamefont{{Jain}}},
  \bibinfo{journal}{\apj} \textbf{\bibinfo{volume}{546}}, \bibinfo{pages}{20}
  (\bibinfo{year}{2001}), \eprint{arXiv:astro-ph/0006319}.

\bibitem[{\citenamefont{{Sato} et~al.}(2010)\citenamefont{{Sato}, {Takada},
  {Hamana}, and {Matsubara}}}]{Satoetal2010}
\bibinfo{author}{\bibfnamefont{M.}~\bibnamefont{{Sato}}},
  \bibinfo{author}{\bibfnamefont{M.}~\bibnamefont{{Takada}}},
  \bibinfo{author}{\bibfnamefont{T.}~\bibnamefont{{Hamana}}}, \bibnamefont{and}
  \bibinfo{author}{\bibfnamefont{T.}~\bibnamefont{{Matsubara}}},
  \bibinfo{journal}{ArXiv e-prints}  (\bibinfo{year}{2010}),
  \eprint{1009.2558}.

\end{thebibliography}
